\documentclass[11pt,letterpaper]{article}
\def\withcolors{0}
\def\withnotes{0}

\usepackage{ccanonne}
\usepackage{palatino}
\usepackage{enumitem}
\usepackage{texmacs}
\newcommand{\arthur}{Arthur\xspace}
\newcommand{\merlin}{Merlin\xspace}
\newcommand{\nimue}{Nimue\xspace}

\title{Verification of Statistical Properties: Redefining the Possible}
\author{{Cl\'ement Canonne}\thanks{The University of Sydney. Email: \email{clement.canonne@sydney.edu.au}.} 
\and {Sam Polgar}\thanks{The University of Sydney. Email: \email{sam.polgar@sydney.edu.au}.} 
\and {Aditya Vikram Singh}\thanks{The University of Sydney. Email: \email{adityavikram.singh@sydney.edu.au}.} 
\and {Sri Aravindan Thyagarajan}\thanks{The University of Sydney. Email: \email{aravind.thyagarajan@sydney.edu.au}.} 
\and {Joy Qiping Yang}\thanks{The University of Sydney. Email: \email{qiping.yang@sydney.edu.au}.} }

\begin{document}

\maketitle

\begin{abstract}
    We revisit the setting of Interactive Proof Systems for Distribution Testing, introduced by Chiesa and Gur (2018), showing that a simple twist on the task requirements may lead to dramatic improvements, allowing verifiers with \emph{constant} sample complexity.
    We define and investigate the multi-prover and zero-knowledge versions of these interactive proof systems, using as flagship example the task of farness verification~--~the ``dual'' version of closeness testing. We hope that our results will inspire others to investigate and analyze the power and limitations of multiple provers for distribution verification.
\end{abstract}


\section{Introduction}

Analyzing (very) large datasets to build accurate models is the workhorse of machine learning and underlies most of the advances in machine learning over the past decade. These datasets are increasingly seen as valuable assets, \eg due to the difficulty in obtaining them (sensitive, regulated, or carefully curated user data), generating them (computation-heavy processes), or trusting them (poisoning attacks). For companies owning such datasets, this leads to a thorny issue: how to convince interested customers that a
dataset has the application-specific statistical properties they need, while revealing as little data as possible?

One avenue forward, inspired by the success of interactive proofs in the functional setting,\footnote{By functional setting, we mean the ``standard'' setting where one seeks to assess whether a fully specified input, encoded as a string, belongs to a given language.} is that of \emph{proofs of proximity for probability distributions}, proposed by Chiesa and Gur~\cite{ChiesaG18}. In this setting, which brings together interactive proofs and (property) distribution testing~\cite{batu2000testing,goldreich1998property}, one sees the ``valuable dataset'' as a probability distribution, and has an interested customer (\arthur) who seeks to leverage interaction with an all-powerful yet untrusted prover (\merlin) to verify whether this distribution $\p$ has a property of interest (or is far from it). This direction of research has recently seen a surge of interest~\cite{HermanR22,HermanR23,HermanR24,HermanR25,GoldreichRothblumG25}, with results showing how to perform a range of fundamental testing tasks~--~for instance, testing any label-invariant property~--~\emph{much} more efficiently, data-wise, than without access to a prover. Yet, in spite of these impressive advances, two key limitations remain: first, these protocols do not ``only'' convince the verifier \arthur that the desired property holds; they actually leak a \emph{lot} of information about \merlin's dataset to him, \emph{enough to learn a the whole frequency profile of the data}. Second, perhaps even more problematic, they all require \arthur to \emph{have a significant number of data points on his own}~--~\ie to himself hold a dataset of size at least $\Omega(\sqrt{\ab})$, where $\ab$ is the domain size, \emph{from the same distribution as \merlin}. This can be seen as deeply unrealistic in many scenarios: How can each potential customer have access to this (sublinear, but still considerable) amount of data from the very dataset they are trying to buy? If they obtain it from some secure, preliminary commitment on the prover's side, is \merlin disclosing large chunks of his data to buyers, one after the other? 

Thus, a natural hope is to remove this strong and unrealistic assumption on \arthur, and seek protocols which only require very little~--~ideally a \emph{constant} number~--~of data point on the verifier's side. (Indeed, as we elaborate later, once this number is low enough, one can indeed justify and envision such a ``secure, preliminary commitment on the prover's side'' to provide \arthur with these very few trusted samples.) Unfortunately, Chiesa and Gur~\cite{ChiesaG18} quashed this hope early on, showing by a very simple and elegant simulation argument that this $\Omega(\sqrt{\ab})$ sample complexity requirement on \arthur's side was necessary even to interactively verify\footnote{Throughout, we will use ``verify'' instead of ``test'' to emphasize the difference with the non-interactive distribution testing setting, and to allow for generalizations of the tasks beyond the usual property testing formulation.} a property as basic as uniformity. This early information-theoretic lower bound seemingly justifies the assumption made in follow-up works, which then focused on getting more bang for this $O(\sqrt{\ab})$ sample complexity buck. Yet, this strong and often unrealistic limitation, even justified by an information-theoretic lower bound, remains a strong and often unrealistic limitation: our work seeks to remove it.
\begin{framed}\centering
    \noindent Is there \emph{any} interesting distribution verification task \arthur can perform with $O(1)$ samples?
\end{framed}
Perhaps unsurprisingly, given the above framing, we argue that the answer is not only \emph{yes}, but also that there are \emph{many} such tasks. Our starting point is the following observation: while uniformity testing is a fundamental distribution testing task, and a building block for many other results~\cite{BatuKR04,CanonneDGR18,Goldreich20}, \emph{it is not in itself a ``goal'' in the distribution verification setting.} That is, it is unlikely that \arthur, a party eager to obtain relevant and hard-to-obtain data for their machine learning or data analysis purposes, hopes to be convinced the dataset on sale is actually ``uniform noise,'' or even from a fixed distribution he could easily sample from by himself~--~and that the seller \merlin's pitch is that he will provide this easily samplable data to anyone willing to buy it.

Instead, one can much more easily motivate situations where \merlin wants to convince \arthur his dataset is \emph{interesting} in some way: some clusters or high-probability records \arthur would be interested in learning more about, or some statistical artifacts worth paying a premium to investigate. In other words, \merlin wants to prove to \arthur the probability distribution he holds is very much \emph{not} uniform. Now, while in the (non-interactive) distribution testing world these questions are equivalent, in the interactive proof setting swapping completeness and soundness might make all the difference: as we show (\cref{theo:uniformity:intro}), there is a very simple protocol allowing to verify \emph{non-uniformity} with a \emph{constant} number of samples! And this is not only the case for (non-)uniformity: as we demonstrate, there exist other natural ``dual'' statistical tasks which allow for constant-sample verification protocols (yet for which non-interactive testing remains hard). Moreover, as we show, this verification can be made without revealing too much to \arthur~--~in a sense which, following the considerable literature on Zero-Knowledge proofs, we make formal in the distribution setting.

\paragraph{The work of Goldreich, Herman, and Rothblum.} Concurrent to and independent from our work, \cite{GoldreichHermanRothblum25} also considers the complement of the closeness promise problem in distribution verification. Their motivation is, however, orthogonal to ours, as they focus on doubly-sublinear single-prover proof systems where both verifier and prover only have sample access to the distributions, and must have sample complexities $o(\sqrt{\ab})$ and $o(\ab)$, respectively~--~that is, the verifier has by themselves too few samples to test, and the prover too few to learn. Goldreich, Herman, and Rothblum provide such a doubly-sublinear protocol, in which one can trade off verifier and prover sample complexities. In contrast, our motivation is the use-case presented above, and our focus is on minimizing the sample complexity of the verifier (Arthur), in a scenario where the prover (Merlin) has already full knowledge of the distribution~--~but seeks to reveals as little about it as possible (the Zero-Knowledge aspect).

\subsection{Our contributions}

\begin{quote}
\emph{``Lower bound arguments do not tell researchers the impossibility results exist. Researchers already know that impossibility results exist. Lower bound arguments tell researchers how the impossibility results can be bypassed.''}\hfill  G.K. Chesterton (mildly paraphrased)
\end{quote}

\noindent The main contribution of this paper is conceptual: namely, revisiting interactive proof systems for distribution testing (\cref{def:ip:distribution}), as introduced in~\cite{ChiesaG18}, to enable a wider range of promise problems~--~and show that many of these, which remain ``hard'' even in the \arthur-\merlin setting, have a natural dual formulation which is not only practically relevant and motivated, \emph{but also becomes easy.}

\begin{theorem}[Verifying farness from uniformity (Informal version of~\cref{theo:nonuniformity})]
    \label{theo:uniformity:intro}
    There exists a simple one-round, communication-efficient protocol with the following guarantees: given a probability distribution $\p$ known to \merlin and provided via sample access to \arthur,
    \begin{itemize}
        \item(Completeness) If $\p$ is far from uniform, then there exists a strategy for \merlin such that \arthur accepts with probability at least $2/3$;
        \item(Soundness) If $\p$ is close to uniform, then no matter what strategy \merlin uses, \arthur accepts with probability at most $1/3$.
    \end{itemize}
    Moreover, \arthur only takes a \emph{constant} number of samples from $\p$.
\end{theorem}
To put this in perspective, recall that this task has sample complexity $\smash{\bigOmega{\frac{\ab}{\log\ab}}}$ in the distribution testing setting (no \merlin)~\cite{ValiantV11}; and $\bigOmega{\sqrt{\ab}}$ samples on \arthur's side in the interactive setting \emph{when swapping completeness and soundness} (via a rather involved protocol~\cite{HermanR22}). \smallskip

\paragraph{Results have consequences.} As a direct corollary of this positive result, we obtain a negative one (\cref{corollary:szk:reduction:hardness}): namely, that an efficient reduction between two tasks, \emph{Entropy Difference} and \emph{Statistical Difference}, which underlies one of the completeness results for the class \textsf{SZK} (Statistical Zero-Knowledge), is impossible in the blackbox-sampling model.

We further show in~\cref{sec:beyond:farness:uniformity} how to build on our farness-from-uniformity verification protocol to systematically obtain protocols to verify farness from a large number of properties: roughly speaking, every class of distributions which can be \emph{learned} with few samples (\cref{theo:farness:property}). This includes, among others, the classes of monotone distributions, $t$-modal and $t$-histogram distributions, log-concave distributions, and Poisson Binomial Distributions.

\paragraph{Defining MIP distribution verification.} Our next contribution is definitional: inspired by the Multi-prover Interactive Proof (MIP) setting~\cite{BenOrGKW88}, and motivated by well-studied distribution testing questions such as closeness testing~\cite{BatuFRS10,ChanDVV13,DiakonikolasK16}, independence~\cite{BatuFFKRW01,AcharyaDK15}, and testing properties of collections of distributions~\cite{LeviRR13,LeviRR12,DiakonikolasK16}, we define the multi-prover version of IP distribution verification, whereby each of $\ell$ distinct non-interacting provers has full knowledge of (only) their own probability distribution $\p_i$, and \arthur, by interacting with these $\ell$ untrusted ``Merlins'', seeks to verify some property of the joint input $(\p_1,\dots,\p_\ell)$. This MIP distribution verification, which is formalized in~\cref{def:mip:distribution}, captures scenarios where distinct (possibly competing) data holders each hold their own dataset, and seek to convince a potential buyer \arthur that it is of interest for him to do so: this justifies the assumption that each ``\merlin'' only has knowledge of \emph{one} distribution $\p_i$, and not the full tuple~--~in contrast to the functional MIP setting of~\cite{BenOrGKW88}.

\paragraph{Farness verification.} With this definition in hand, we turn to a second verification task: the dual of closeness testing, which we term \emph{farness verification.} Here, two provers~--~\merlin and \nimue~--~each have full knowledge of only their own probability distribution, $\p$ and $\q$, and seek to convince \arthur that $\p$ and $\q$ are statistically \emph{far} from each other. One could see this arising from a scenario where \arthur is willing to get diverse, complementary data from different sources, to have a more representative training set: and so two datasets coming from the same distribution are of little appeal, while two datasets statistically far from each other will meet his goal. It is known that, in the non-interactive testing setting, $\Omega(\ab^{2/3})$ samples are required from at least one of the two distributions~\cite{BatuFRS10,Valiant11, ChanDVV13,DiakonikolasK16} for this farness testing task (which is then equivalent to closeness testing); and that by the same simulation argument as for uniformity, testing closeness in the interactive proof setting necessitates $\Omega(\sqrt{\ab})$ samples from \arthur~\cite{ChiesaG18}.  

\begin{theorem}[Verifying farness (Informal version of~\cref{theo:farness})]
    \label{theo:farness:intro}
    There exists a simple one-round, communication-efficient protocol with the following guarantees: given two probability distributions $\p$ and $\q$, with $\p$ known to \merlin and $\q$ to \nimue, respectively, and provided via sample access to \arthur,
    \begin{itemize}
        \item(Completeness) If $\p$ is far from $\q$, then there exists a strategy for \merlin and $\nimue$ such that \arthur accepts with probability at least $2/3$;
        \item(Soundness) If $\p=\q$, then no matter what strategy \merlin and \nimue use, \arthur accepts with probability at most $1/2$.
    \end{itemize}
    Moreover, \arthur only takes a \emph{constant} number of samples from $\p$ and $\q$.
\end{theorem}
Along the way, we also establish some variants of the above theorem, allowing for stronger soundness guarantees under more stringent assumptions on $\p,\q$. We further show in~\cref{sec:farness:clusterability} how this farness verification protocol can be used as a building block to obtain a verification protocol for \emph{non-clusterability} of distributions (\cref{theo:clusterability}).

\paragraph{Zero-Knowledge.} Having (re)defined IP distribution verification, and introduced its multi-prover generalization, it is natural to ask how to address this first ``key limitation'' alluded to earlier: that is, the fact that all existing IP distribution testing protocols in the literature leak a considerable amount of information from \merlin to \arthur. Our next contribution is to define the natural analogue of \emph{Zero-Knowledge proofs} to the (single- and multi-prover) distribution verification setting (\cref{def:mip:distribution:zk}), both statistical and perfect. We also show that these definitions are \emph{achievable}: specifically, we show in~\cref{theo:nonuniformity} that the non-uniformity verification protocol of~\cref{theo:uniformity:intro} satisfies our Statistical Zero-Knowledge definition, and, later on (\cref{lemma:farness:zk:perfectzk}), that a variant of our farness verification protocol (under stronger assumptions) achieves Perfect Zero-Knowledge. 

In view of our motivation for the multi-prover setting, we also introduce a new, weaker notion of Zero-Knowledge proof, which is easier to attain yet does capture an important desirable property: that is, ``any given prover should not learn much about the \emph{other} provers' data.'' This definition, which we formalize in~\cref{def:mip:distribution:opozk}, can be interpreted as follows: given an MIP distribution verification protocol $\Pi$  with $\ell$ provers, for every $i\in[\ell]$, there exists an algorithm which simulates the view of prover $M_i$ from $\Pi$ in the completeness case (\ie where everyone is honest), yet only is given input $\p_i$ and the same sample access to $\p_1,\dots, \p_\ell$ as \arthur. In other words, this definition ensures that whatever $M_i$ learns about $\p_1,\dots, \p_{i-1},\p_{i+1},\dots,\p_\ell$ when running a (valid) instance of the protocol $\Pi$ on a ``\yes-input'', it could have learned in an ``isolated execution'' involving only himself and \arthur, pretending to run the protocol $\Pi$ with \emph{none of the other provers}. We show that in~\cref{theo:farness} that the non-uniformity verification protocol of~\cref{theo:farness:intro} achieves this ``One-Merlin-Out'' Statistical Zero-Knowledge definition.

\paragraph{Repetition theorems.} 
The completeness and soundness parameters set above to $2/3$ and $1/3$, respectively, may seem arbitrary: as ``standard'' in randomized computation and the interactive proof literature, one expects to be able to amplify them \emph{via} repetition. In~\cref{sec:repetition}, we show that this is indeed the case, establishing the analogues of repetition theorems from the (function) IP and ZK literature for the distribution verification setting. In particular, the completeness parameter can be made exponentially close to $1$ and the soundness parameter exponentially close to $0$, thanks to guarantees provided by sequential repetition theorem (\cref{theo:sequential:repetition:mip}). Whether the same can be achieved using \emph{parallel} repetition is something we leave as a future direction.

\paragraph{Multi-prover lower bounds for uniformity testing.}
\new{We note that, in contrast to the functional MIP setting, allowing for multiple \merlin's holding the same distribution does not systematically provide more power, and in particular does not enable one to bypass the $\Omega(\sqrt{\ab})$ sample complexity lower bound for verifying uniformity. Indeed, the simulation argument from~\cite[Observation~3.10]{ChiesaG18} carries over from the single-prover case, as \arthur could otherwise simulate the honest provers strategies (answering consistently with the single, fixed \yes-instance $\uniformOn{\ab}$), obtaining from interactive protocol a valid testing protocol with too-good-to-be-true sample complexity.}

\subsection{Definitions: (Multi)-Prover distribution verification, and capturing Zero-Knowledge}

We begin by recalling (and suitably extending) the definition of Chiesa and Gur:

\begin{definition}[{IP distribution verification (Slight generalization of~\cite[Definition~5.2]{ChiesaG18}})]
    \label{def:ip:distribution}
    Given two disjoint properties $\property_0,\property_1 \subseteq \distribs{\domain}$ over a known domain $\domain$ of size $\ab \eqdef \abs{\domain}$,\footnote{As common in the property testing literature, we slightly abuse notation and refer to a property $\property$ corresponding to a fixed size parameter $\ab$, instead of an (infinite) family $\property = \{\property_{\ab}\}_{\ab\in \N}$.} an \emph{interactive proof system for distribution verification} $\Pi$ of properties $\property_0,\property_1 \subseteq \distribs{\domain}$ consists of an algorithm $A$ which interactively exchanges messages with a prover $M$ such that at the end of the interaction the following two conditions hold.
        \begin{itemize}
            \item(Completeness) for all $\p \in \property_0$, there exists a prover strategy $M$ such that 
            \[
                \probaOf{\Pi(M(\p),A^{\p}) =  \accept } \geq 2/3\,.
            \]
            \item(Soundness) for all $\p \in \property_1$, for all prover strategy $M^\ast$, 
            \[
                \probaOf{\Pi(M^\ast(\p),A^{\p}) =  \accept } \leq 1/3\,,
            \]
        \end{itemize}
        where $A^{\p}$ indicates that $A$ has sample access to $\p$. The \emph{sample complexity} of $\Pi$ is the (worst-case) number of samples $A$ draws from the distribution, the
\emph{communication complexity} of $\Pi$ is the (worst-case) total number of bits exchanged between the parties, and
the \emph{round complexity} of $\Pi$ is the (worst case) number of rounds of interaction, where each round consists of
a message from one party to the other and its reply. We say that $\Pi$ is \emph{efficient} when the algorithm $A$ is running in time polynomial in $\ab$ and its sample complexity.
\end{definition}
The above choice of $2/3$ for the probability of success of the proof system may seem arbitrary, and~--~as is the case in the standard non-interactive setting~--~easy to amplify by standard arguments. However, amplification by repetition theorems are much less trivial in the interactive setting (see, \eg~\cite[Appendix~C.1]{Goldreich98} and~\cite{Goldreich25}). In~\cref{sec:repetition}, we expand on this, checking that the argument for parallel and 
sequential repetition in the ``functional'' (single-prover) IP setting do carry over to the (single-prover) distribution verification IP one.

\paragraph{``Functional'' multi-prover interactive protocols (MIP)~\cite{BenOrGKW88}.} In the functional setting, an $\ell$-prover proof system for a language $L$ is a protocol $\Pi$ between $\ell$ (computationally unbounded) provers $(M_1, \ldots, M_\ell)$ and (probabilistic polynomial-time) verifier $A$ such that (1) $M_1,\ldots,M_\ell$, $A$ are all given the input $x$, but for every $i \neq j$, $M_i$ does not see the messages between $M_j$ and $A$; (2) $M_1,\dots,M_\ell$ all have access to a read-only shared uniformly random string $r\in\{0,1\}^\ast$, independent of the input $x$; and, (3) $\Pi$ satisfies the following correctness conditions:
        \begin{itemize}
            \item(Completeness) for all $x \in L$, $\probaOf{\Pi(M_1,\ldots,M_\ell,A)(x) =  \accept } \geq 2/3$;
            \item(Soundness) for all $x \notin L$, for all $P'_1,\ldots,P'_\ell$, $\probaOf{\Pi(P'_1,\ldots,P'_\ell,A)(x) =  \accept } \leq 1/3$;
        \end{itemize}
where the probability is taken over $r$ and the provers and verifier's own private randomness. Importantly, in contrast to the single-prover case, while sequential repetition theorems are known to increase the probability of success to be arbitrarily close to 1, success probability amplification cannot be done in parallel in the blackbox way (see, \eg~\cite[Section~6]{FortnowRS94}). 

\paragraph{Motivating the multi-prover model.} In our case, for closeness testing:
\begin{enumerate}[itemsep=1pt]
    \item The input is a pair of distributions $(\p,\q)$.
    \item There are two powerful provers $M$ and $N$, and a sample-bounded verifier $A$.
    \item $M$ has access to $\p$ (but not $\q$), $N$ has access to $\q$ (but not $\p$), and $A$ has (sample) access to $(\p,\q)$. 
    \item Completeness: If $\totalvardist{\p}{\q}\geq 9/10$, then $\probaOf{\Pi(M,N,A) =  \accept } \geq 2/3$. 
    \item Soundness: If $\totalvardist{\p}{\q} \leq 1/10$, then for all $M',N'$ $\probaOf{\Pi(M',N',A) =  \accept }  \leq 1/3$.
\end{enumerate}
We will in the remainder of this paper mostly focus on either the single-prover setting (between \arthur and \merlin, abbreviated as $A$ and $M$) or the 2-prover setting (involving \arthur, \merlin, and \nimue, abbreviated as $A$, $M$, and $N$). However, for the sake of generality, and as we believe this extension to be of interest, we define the $\ell$-prover case, for arbitrary $\ell\geq 2$.


\begin{definition}[MIP distribution verification]
    \label{def:mip:distribution}
    Given two disjoint properties $\property_0,\property_1 \subseteq \distribs{\domain_1\times\dots\times\domain_\ell}$ over known domains $\domain_1,\dots,\domain_\ell$ of size $\ab_1,\dots,\ab_\ell$, a \emph{multi-prover interactive proof system for distribution verification} $\Pi$ of properties $\property_0,\property_1$ consists of an algorithm $A$ which interactively exchanges messages with provers $M_1,\dots,M_\ell$ (which share a read-only uniformly random string $r\in\{0,1\}^\ast$, independent of everything else, but do not communicate with each other) such that at the end of the interaction the following two conditions hold.
        \begin{itemize}
            \item(Completeness) for all $(\p_1,\dots,\p_\ell) \in \property_0$, there exists a prover strategy $(M_1,\dots,M_\ell)$ such that 
            \[
                \probaOf{\Pi(M_1(\p_1),\dots,M_\ell(\p_\ell),A^{\p_1,\dots,\p_\ell}) =  \accept } \geq 2/3\,.
            \]
            \item(Soundness) for all $(\p_1,\dots,\p_\ell) \in \property_1$, for every prover strategy $(M^\ast_1,\dots,M^\ast_\ell)$, 
            \[
                \probaOf{\Pi(M^\ast_1(\p_1),\dots,M^\ast_\ell(\p_\ell),A^{\p_1,\dots,\p_\ell}) =  \accept } \leq 1/3\,,
            \]
        \end{itemize}
        where $A^{\p_1,\dots,\p_\ell}$ indicates that $A$ has (independent) sample access to $\p_1,\dots,\p_\ell$. The \emph{sample complexity} of $\Pi$ is the (worst-case) total number of samples $A$ draws from the distributions, the
\emph{communication complexity} of $\Pi$ is the (worst-case) total number of bits exchanged between the parties, and
the \emph{round complexity} of $\Pi$ is the (worst case) number of rounds of interaction, where each round consists of
a message from one party to the other and its reply. We say that $\Pi$ is \emph{efficient} when the algorithm $A$ is running in time polynomial in $\ab_1,\dots,\ab_\ell$ and its sample complexity.
\end{definition}

\begin{remark} In contrast with the ``functional'' MIP, here $M_1,M_2\dots, M_\ell$ only have \emph{partial} access to the input $(\p_1,\p_2,\dots,\p_\ell)$. One the one hand, one can hope this weaker access to the input allows us to obtain soundness guarantees with a \emph{constant} number of samples; on the other, it makes proving completeness more challenging, as even the ``all-powerful provers'' only have limited information. This also makes for a richer landscape when it comes to defining notions of zero knowledge since we can impose zero knowledge on the provers $M_1,\dots, M_\ell$ too: as we will now see.
\end{remark}


\paragraph{Defining Zero Knowledge for MIP distribution verifiers.} We now introduce two different notions of Zero-Knowledge for MIP distribution verification, starting with the direct analogue of the (Statistical) Zero-Knowledge in the functional setting.
\begin{figure}
    \centering
    \includegraphics[width=0.8\linewidth]{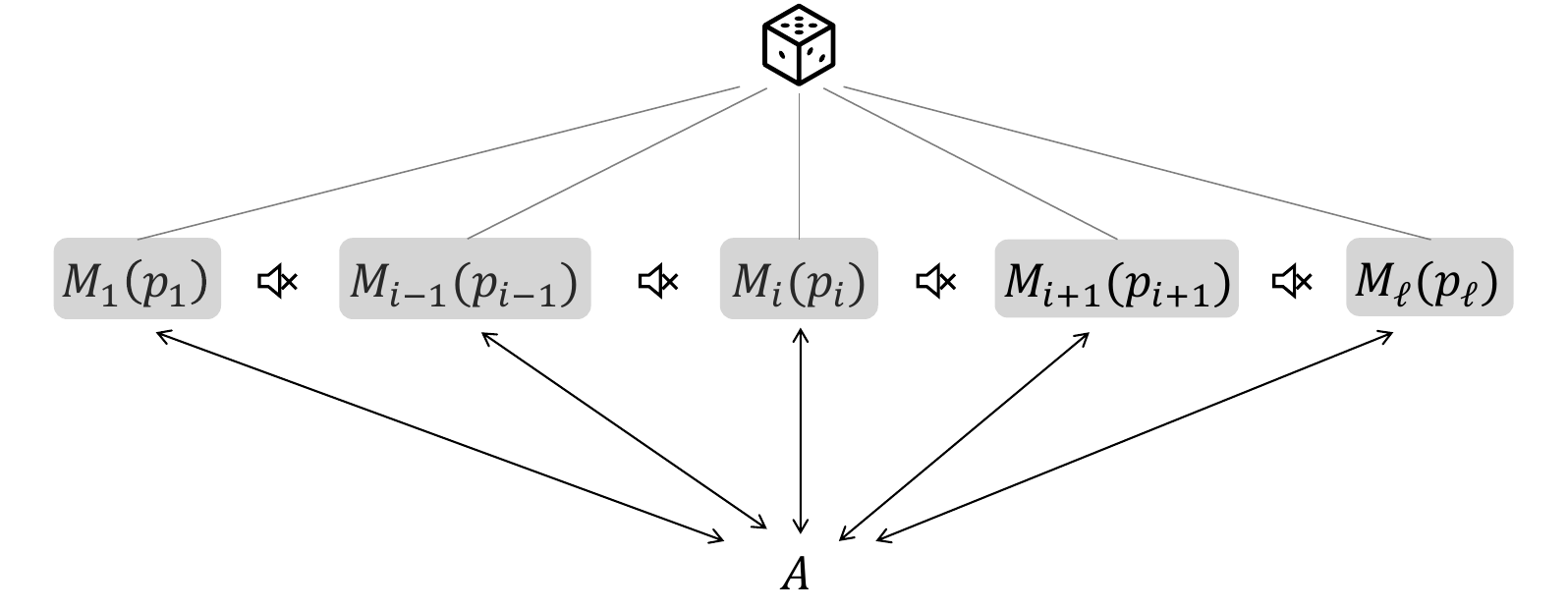}
    \caption{The MIP distribution verification setting. The $\ell$ provers (``Merlins'') share a read-only random tape, independent of their respective inputs, but are not allowed to communicate otherwise. The $i$-th Merlin is given full knowledge of their part of the input, namely, distribution $\p_i$, and does not know or have access to $\p_j$ for $j\neq i$.}
    \label{fig:manymerlins}
\end{figure}

\begin{definition}[ZK for MIP distribution verification]
    \label{def:mip:distribution:zk}
    An MIP proof system $\Pi=(M_1,\dots, M_\ell, A)$ for distribution verification properties $\property_0,\property_1$ is said to be \emph{(honest-verifier) statistical zero-knowledge}  if there is a randomized algorithm $S$ and a non-increasing function $\errprob\colon\N\to[0,1]$ such that, for every $(\p_1,\dots,\p_\ell)\in\property_0$ and all $t\in\N$,
    \[
        \totalvardist{S^{\p_1,\dots,\p_\ell}(t)}{\dotprod{A^{\p_1,\dots,\p_\ell}}{M_1(\p_1),\dots, M_\ell(\p_\ell)}_{A}(t)} \leq \errprob(t)\,,
    \]
    where $S$ has the same sample complexity as $\Pi$ (up to constant factors), and $\dotprod{A}{M_1,\dots,M_\ell}$ denotes the \emph{view} of the interaction (\ie the transcript) between all parties, from \arthur's perspective (including its own randomness and samples from $\p_1,\dots,\p_\ell$). When $\errprob=0$ (constant function), we say that $\Pi$ is (honest-verifier) \emph{perfect} zero-knowledge. 
\end{definition}

As in the functional case, we emphasize that the simulator is only required to work in the completeness case (when the input belongs to $\property_0$). However, there is an important difference with the functional case: namely, that we do not enforce computational efficiency, but instead require that the simulator only be given \emph{sample access} to the input distributions (\ie the same access as \arthur) \emph{and} uses a comparable number of samples.\footnote{Note that we allows some constant-factor slack here, to allow sufficient robustness in the definition.} This sample complexity requirement is key, as otherwise the simulator would be allowed to draw a number of samples sufficient to, \eg learn the input distributions.

Having defined ZK, which captures that \arthur should not learn much from the interaction, we turn to the weaker, new notion we discussed earlier: a notion of Zero-Knowledge which captures that any given \emph{prover} should not learn much about the \emph{other} provers' inputs. 

\begin{definition}[HIPOZK for MIP distribution verification]
    \label{def:mip:distribution:opozk}
    An MIP proof system $\Pi=(M_1,\dots, M_\ell, A)$ for distribution verification properties $\property_0,\property_1$ is said to be \emph{Statistical Honest Interactive Prover Only (HIPO) zero-knowledge} if there is a randomized algorithm $S$ and a non-increasing function $\errprob\colon\N\to[0,1]$ such that, for every $(\p_1,\dots,\p_\ell)\in\property_0$, every $i\in [\ell]$, and all $t\in\N$,
    \[
        \totalvardist{S^{\p_1,\dots,\p_\ell}(\p_i,t)}{\dotprod{A^{\p_1,\dots,\p_\ell}}{M_1(\p_1),\dots, M_\ell(\p_\ell)}_{M_i}(t)} \leq \errprob(t)\,,
    \]
    where $S$ has the same sample complexity as $\Pi$ and $\dotprod{A^{\p_1,\dots,\p_\ell}}{(M_1(\p_1),\dots, M_\ell(\p_\ell))}_{M_i}(t)$ denotes view of the interaction between all parties, from the $i$-th prover perspective. When $\errprob=0$, we say that $\Pi$ is \emph{perfect} HIPOZK.
\end{definition}

\begin{remark}[Generalizing HIPOZK to arbitrary subset of Merlins]
Instead of simulating the interaction between all parties from the $i$-th prover's perspective as in \cref{def:mip:distribution:opozk}, we can also define zero knowledge for a fixed subset of provers, i.e., among that subset of provers, they do not learn more than the number of samples they could see from the protocol. This would capture the setting where a few provers (companies) can collude to learn secrets of other provers (companies).
\end{remark}

\new{
\begin{remark}[Comparison to secure multiparty computation]
    Our notion of HIPOZK, like the simulation paradigm of \cite{Lindell20}, seeks to capture that a party learns nothing beyond what it is entitled to (and that its simulator view must be computed from this). Classic zero knowledge defines this for the verifier view and HIPOZK defines it for a prover view. One can see the ``ideal model'' described in~\cite[Section~6.2]{Lindell20} as related to our notion of HIPOZK, for the setting of multiparty computation: however, there are a few crucial difference, which we believe call for our definition. Some are technical (computational v. statistical indistinguishability, access to samples for the trusted party, v. access to the inputs, etc.); perhaps more importantly, there is a conceptual difference, in that our notion captures the protocol itself, while the ideal model described in \cite{Lindell20} is a proof technique to define the simulation.
\end{remark}
}


\section{Related Work}
\emph{Property testing}, introduced in \cite{RubinfeldS96, goldreich1998property}, is the study of ultra-efficient randomized algorithms for approximate decision-making on \emph{unknown} massive objects. These algorithms, called testers, are given \emph{local query access} to an oracle representing the object, \eg as a function $f$. With only a number of queries sublinear in the size $n$ of the object's representation, the tester must decide whether $f$ is in the property $\property$ of interest, or $\dst$-far (\ie at distance at least $\dst$) from every $g\in \property$, where the distance depends on the type of objects considered (\eg Hamming distance for functions). See \cite{ Fischer01, fischer2004art, Ron08, Goldreich10, Goldreich17} for introductions. 

In the \emph{distribution testing} setting~\cite{goldreich1998property,batu2000testing}, the ``objects'' are probability distributions over a discrete domain of size $\ab$, the type of queries is (typically) \iid sample access to the distribution, and the measure of distance is (again, typically) the total variation distance. Uniformity~\cite{GoldreichRon99, Paninski08}, identity~\cite{BatuFFKRW01,AcharyaDK15}, and closeness~\cite{BatuFRSW00, AcharyaDJOPS12, ChanDVV13, BatuFRSW13,DiakonikolasK16} testing were some of the earliest and fundamental tasks considered, and all require a sampling complexity polynomial in the domain size ($\Theta(\sqrt{\ab}/\dst^2)$ for uniformity and identity, and $\Theta(\max(\ab^{2/3}/\dst^{4/3},\sqrt{\ab}/\dst^2))$ for closeness, respectively).

A more challenging task in property testing is that of \emph{tolerant} testing~\cite{ParnasRR06}, where one seeks to distinguish objects $\dst_1$-close to the property from those $\dst_2$-far from it. In the distribution testing setting, this is a much more challenging task, with tolerant uniformity testing requiring $\Theta(\ab/\log\ab)$ samples, for instance~\cite{ValiantV11,CanonneJKL22}. Tolerant testing has been thoroughly investigated, especially for the broad class of \emph{symmetric (label-invariant) properties} (that is, properties closed under relabeling of the domain), such as entropy and support size~\cite{BatuDKR05, GuhaMV09, ValiantV11, ValiantV17,RaskhodnikovaRSS09,WuY16}.

The above are concerned with the non-interactive setting, in which the testing algorithm has access to the ``object'' -- but no external help.  Rothblum, Vadhan, and Wigderson introduce \emph{Interactive Proofs of Proximity (IPP)} \cite{RothblumVW13} in the functional setting following work from \cite{ErgunKR04}, where an unbounded and possibly adaptively cheating prover receives $x$, interacts with a sublinear-time verifier to decide whether $x \in L$ or is $\dst$-far from $L$. They answer the question of whether a proof system can improve complexity bounds for property testing by showing exponential improvements over the standard property testing model for specific properties.

\cite{BermanRV18}, motivated by the question of whether a prover convince a verifier of correct computation if the verifier only reads a few bits of input, without revealing additional non-local information about the input, defines the Zero-Knowledge Proof of Proximity (ZKPP) setting, where an unbounded and potentially malicious prover seeks to convince a sublinear-time verifier that the unknown object is $\dst$-close to a property, without revealing \emph{non-local information}. Unlike classical ZK \cite{GoldwasserMR89, GoldreichMW91} where a verifier sees the input $x$ and learns whether $x \in L$ (without revealing anything else), ZKPP restricts the verifier to oracle access to the object (e.g., function $f$ or graph $G$) and reveals no information beyond what the queries.

Combining the above two then-disjoint lines of inquiry, Chiesa and Gur~\cite{ChiesaG18} introduced the notion of proof systems for properties of distributions, \emph{Proofs of Proximity for Distribution Testing}. The question there was to understand whether an untrusted prover with complete knowledge (or, alternatively, unbounded sample access) to a distribution can prove membership in a property to a verifier who only has sample access~--~using significantly fewer samples than required for testing the property. Chiesa and Gur initiated a systematic treatment of the question, introducing variants of the models (MA, AM, and IP distribution testers), and investigated the power and advantages conferred by public- and private-coin protocols, and interaction between prover and verifier. A line of follow-up works by Herman and Rothblum~\cite{HermanR22, HermanR23, HermanR24, HermanR25} improved efficiency first for symmetric properties, then for general properties; however, their results all require at least $\Omega(\sqrt{\ab})$ samples on the verifier's side. \cite{GoldreichRothblumG25} and \cite{HermanR25} specifically investigate whether (and for which properties) both verifier and prover could have sample complexity significantly smaller than their ``corresponding task''~--~namely, smaller than the testing sample complexity for the verifier, and smaller than the learning complexity for the prover.

Finally, the (functional) setting of \emph{multiple} provers was introduced in~\cite{BenOrGKW88}, and investigated (among others) in~\cite{FortnowRS94}. \cite{CanettiRR11, CanettiRR13, GoldwasserKR15, CanettiLW25} then specifically consider this multiprover setting in the context of delegated computation.

We finally mention two other relevant works. First, that of Goldwasser, Rothblum, Shafer, and Yehudayoff~\cite{GoldwasserRSY21}, which introduces a setting analogous to that of Chiesa and Gur~\cite{ChiesaG18}, but for verifying \emph{PAC learning} instead of testing. Second, that of Tell~\cite{Tell16}, which, in the (noninteractive, testing-only) setting, studies the \emph{dual problems} of property testing in depth: namely, requiring that the tester accept when the object is $\dst$-far from a property $\property$ and reject if it is $\dst$-far from the set of objects that are $\dst$-far from $\property$.  As a special case of~\cite[Theorem 1.9]{Tell16}, it is shown that the dual of testing uniformity (testing farness from uniformity) is the same as testing uniformity (when support size is sufficiently larger than $1/\dst$). Moreover, in general, when testing without provers, dual problems are at least as hard to test as the original problems~\cite[Observation 1.4]{Tell16}).

\subsection{Future Directions: the Holy Grail (\arthur without samples)}

Our results leave open how far Arthur’s \emph{sample} complexity can be pushed while still certifying interesting distributional properties. We highlight two complementary relaxations.

\paragraph{Hidden modes with game-theoretic incentives.}
Recent work shows that some tasks deemed impossible or unfair in the standard cryptographic model can be recovered once the adversary is associated with utility functions, and it wants to maximise its utility~\cite{Cleve86,AndrychowiczSP14,BentovKumaresan14,Back14,BaumEtAl20,chung2018game,wu2022complete,thyagarajan2024game}. Motivated by this, we could couple \emph{hidden modes} with incentives: Arthur sometimes has \emph{no} samples, but crafts challenges whose distribution is indistinguishable from those in rounds where he does have samples, so Merlin cannot tell the mode. Under a utility that rewards Arthur's acceptance and imposes a loss or $0$ utility for Arthur's rejection, Merlin maximizes utility by responding exactly as if Arthur had samples in every round. This preserves soundness and lets a small fraction of with-sample rounds lift completeness, thereby lowering Arthur’s \emph{expected} sample complexity. The main tasks are to (i) define and enforce mode-indistinguishability at the transcript level, (ii) calibrate the with-sample fraction and payoff parameters to meet a target $\delta$-soundness and completeness, and (iii) prove robustness to collusion/adaptivity without leaking the mode.

\paragraph{Cryptographic assistance.}
If we move away from the information-theoretic setting to a computational one where we allow computational or setup assumptions, we can let Merlin (or Merlins) provide succinct, verifiable certificates (\eg commitments plus short arguments) that shift cost from Arthur's sample complexity to communication. This points to protocols where Arthur, for instance, uses only a constant or amortized subconstant number of fresh samples to spot-audit a proof of ``non-uniformity'' computed over the full dataset. One has to characterize tradeoffs between Arthur’s samples, argument length and verification time, and whether reusable certificates can push per-instance sampling below current constants.

Along the way, we also state a few open problems of interest, starting with a very natural one.
\begin{openquestion}
    Which other statistical tasks admit a dual version which can be verified with only a constant number of samples?
\end{openquestion}
\new{Moreover, all our ZK guarantees hold are \emph{honest verifier} guarantees. This raises the question of whether they can be made secure even against a dishonest \arthur:
\begin{openquestion}
    Which protocols for distribution verification can be made (malicious-verifier) ZK? Is there, as in the functional setting, a general transformation from honest- to malicious-verifier?
\end{openquestion}}

\noindent\textbf{AI Disclosure:} No AI tool was used in the preparation of this manuscript, in any capacity whatsoever. We used coffee to assist with staying awake and focused during the writing of the proofs. The tool materially affected our brains, in various ways, mostly positive. 

\section{Preliminaries}
We will use throughout the standard $\bigO{\cdot}$ notation, and write $\distribs{\ab}$ to denote the set of discrete distributions supported on a domain $\domain$ of size $\ab$ (without loss of generality, assume that the domain is $[\ab]=\{1,2,\dots,\ab\}$).
For two discrete distributions $\p \in \distribs{\ab}$ and $\q \in \distribs{\ab}$, the total variation distance is defined as $\totalvardist{\p}{\q} = \sup_{S\subseteq \domain} \p(S)-\q(S) = \frac{1}{2}  \sum_{x \in \errprob (\ab)} | \p (x) - \q (x)|$.

We now recall two results we will rely upon, the Polarization Lemma and the Pearson--Neyman Lemma. While originally written for distributions sampleable from (explicitly given) \emph{circuits}, one can check by inspection of the proof that the former also applies to distributions provided \emph{via} samples only. For completeness, we restate a proof sketch of the Polarization Lemma in Appendix~\ref{sec:appendix}.

\begin{lemma}[{Polarization Lemma~\cite[Lemma~3.1.12]{Vadhan99}}]
\label{lemma:polarization}
Let $\alpha,\beta\in[0,1]$ be any two constants such that $\alpha^2>\beta$. 
There exists a polynomial-time computable mapping which, given sample access to two discrete probability distributions $\p,\q$ (over a domain of size $\ab$) and parameter $\errprob\in(0,1]$, provides sample access to two discrete probability distributions $\p',\q'$ (over a domain of size $\ab'$) such that
\begin{itemize}
    \item If $\totalvardist{\p}{\q} \geq \alpha$, then $\totalvardist{\p'}{\q'} \geq 1-\errprob$;
    \item If $\totalvardist{\p}{\q} \leq \beta$, then $\totalvardist{\p'}{\q'} \leq \errprob$;
\end{itemize}
and $\ab' = \ab^{s}$, where $s = (\log (1 / \errprob))^{\frac{2 \log (1 / \alpha)}{\log (\alpha^2 /
       \beta)}+ O(\log (1 / \alpha))}$. Moreover, one can efficiently sample from $\p'$ or $\q'$ given $s$ samples from $\p$ and $\q$.
\end{lemma}

\begin{lemma}[{Pearson-Neyman Lemma (Folklore; see, \eg~\cite[Lemma 1.4]{Canonne22})}]\label{lemma:pearson-neyman}
  Any (possibly randomized) statistical test which distinguishes between $\p$
  and $\q$ from a single sample must have Type I (false positive) and Type-II
  (false negative) errors satisfying
  \[
    \textrm{Type I} + \textrm{Type II} \geq 1 - d_{\tmop{TV}} (\p, \q)\,.
\]
\end{lemma}

\section{Verifying non-uniformity}
In this section, we prove our main theorem in the single-prover setting, before discussing its consequences, extensions, and applications.
\begin{theorem}
    \label{theo:nonuniformity}
    There exists a one-round private-coin protocol $\Pi=\Pi(A,M)$ with the following guarantees: on input $\ab \geq 1$, and constants $\errprob\in(0,1]$, $\dst_1,\dst_2\in[0,1]$ with $\dst_1 < \dst_2^2$; and sample access to an unknown probability distribution $\p\in\distribs{\ab}$,
    \begin{itemize}
        \item(Completeness) If $\totalvardist{\p}{\uniformOn{\ab}} \geq \dst_2$, then $\probaOf{\Pi(A,M) = \accept } \geq 1-\errprob$;
        \item(Soundness) If $\totalvardist{\p}{\uniformOn{\ab}} \leq \dst_1$, then, for all $M^\ast$, $\probaOf{\Pi(A,M^\ast) = \accept } \leq \errprob$;.
    \end{itemize}
    Moreover, $A$ takes $\Paren{\log(1/\errprob)}^{\bigO{1/\log(\dst_1^2/\dst_2)}} = \bigO{1}$ samples, and the protocol is (honest-verifier) Statistical Zero-Knowledge.
\end{theorem}

\begin{algorithm}[htbp]
    \begin{algorithmic}[1]
        \Require Sample access to $\p$ (\arthur) over domain $\domain$, full knowledge of $\p$ (\merlin), full knowledge of a reference distribution $\p^\ast$ (\merlin), either known or samplable by \arthur.
        \LineComment{\textbf{\arthur}}
        \State Chooses $b \samples \{0,1\}$, 
        \State Sample $x_0\sim \p^\ast$ and $x_1\sim \p$
        \State Send $x_b$ to \merlin
        \LineComment{\textbf{\merlin}}
        \State Upon receiving $x_b$, must guess $\hat{b}\in\{0,1\}$
        \State Send $\hat{b}$ to \arthur
        \LineComment{\textbf{\arthur}}
        \State \Return \accept if $\hat{b}=b$.
    \end{algorithmic}
    \caption{Basic protocol for verifying non-identity}\label{algo:nonuniformity:basic}
\end{algorithm}

The proof of this theorem will rely on the following primitive to verify non-uniformity ``with small advantage.'' Both for the sake of generality and in view of building upon this primitive later on, we provide a more general statement, allowing to verify non-\emph{identity} to a reference distribution $\p^\ast$. (Throughout, the reader may think of $\p^\ast$ as $\uniformOn{\ab}$.)
\begin{lemma}
    \label{lemma:nonuniformity:basic}
    There exists a one-round private-coin protocol $(A,M)$,~\cref{algo:nonuniformity:basic}, with the following guarantees: on input $\ab \geq 1$, $\dst_1,\dst_2\in[0,1]$ with $\dst_1 < \dst_2$, sample access to an unknown probability distribution $\p\in\distribs{\ab}$, and given a reference distribution $\p^\ast \in\distribs{\ab}$,
    \begin{itemize}
        \item(Completeness) If $\totalvardist{\p}{\p^\ast} \geq \dst_2$, then $\probaOf{\Pi(A,M) = \accept } \geq \frac{1}{2}+\frac{\dst_2}{2}$;
        \item(Soundness) If $\totalvardist{\p}{\p^\ast} \leq \dst_1$, then, for all $M^\ast$, $\probaOf{\Pi(A,M^\ast) = \accept } \leq \frac{1}{2}+\frac{\dst_1}{2}$;.
    \end{itemize}
    Moreover, $A$ takes \emph{one} sample, and the total communication is $\bigO{\log\ab}$ bits.
\end{lemma}
\begin{proof}
The lemma will directly follow from the next two claims, which, respectively, establish completeness and soundness of the protocol. Its sample and communication complexity are immediate.
\begin{claim}[Completeness]
    If $\totalvardist{\p}{\p^\ast} \geq \dst_2$, then there exists a strategy $M$ for \merlin such that the protocol of~\cref{algo:nonuniformity:basic} returns \accept with probability at least $\frac{1+\dst_2}{2}$.
\end{claim}
\begin{proof}
    Consider the so-called Scheff\'e set 
    \[
        S_\p \eqdef \setOfSuchThat{x\in \domain}{ \p(x) \geq \p^\ast(x)}
    \]
    which \merlin can compute explicitly, having full knowledge of the distribution $\p$ (and $\p^\ast$). \merlin's strategy $M$ is then simply to return the bit $\hat{b} \eqdef \indicSet{S_\p}(x_b)$, \ie the indicator of whether the sample sent by Arthur falls in this Scheff\'e set.
    
    By standard facts about total variation distance, $\p(S_\p) - \p^\ast(S_\p) = \totalvardist{\p}{\p^\ast} \geq \dst_2$; as a result,
    \begin{align*}
        \probaOf{\hat{b}=b} 
        &= \probaCond{x_0 \notin S_\p}{b=0}\probaOf{b=0} + \probaCond{x_1 \in S_\p}{b=1}\probaOf{b=1}\\
        &= \frac{1}{2}\Paren{\p^\ast(\overline{S_\p}) + \p(S_\p)}\\
        &= \frac{1}{2}\Paren{1- \p^\ast(S_\p) + \p(S_\p)}\\
        &\geq \frac{1}{2}\Paren{1+\dst_2}\,,
    \end{align*}
    as claimed.
\end{proof}

\begin{claim}[Soundness]\label{claim:basicsoundness}
    If $\totalvardist{\p}{\p^\ast} \leq \dst_1$, then there is no strategy $M$ for \merlin such that the protocol of~\cref{algo:nonuniformity:basic} returns \accept with probability greater than $\frac{1+\dst_1}{2}$.
\end{claim}
\begin{proof}
    This follows from the characterization of total variation distance as best possible advantage for distinguishing two fixed distributions given a single observation. Namely, any possibly randomized strategy of \merlin to guess $b$ can be seen as a distribution over deterministic strategies: each such strategy $M^\ast$ corresponds to outputting the indicator of a subset $S_{M^\ast} \subseteq \domain$: but the total variation distance is $\totalvardist{\p}{\p^\ast} = \sup_{S\subseteq \domain} ( \p(S) - \p^\ast(S) ) \geq \p(S_{M^\ast})-\p^\ast(S_{M^\ast})$; and so, as in the proof of the previous claim,
    \begin{align*}
        \probaOf{\hat{b}=b} 
        &= \frac{1}{2}\Paren{1- \p^\ast(S_{M^\ast}) + \p(S_{M^\ast})}
        \leq \frac{1}{2}\Paren{1+\dst_1}\,,
    \end{align*}
    proving the claim for deterministic strategies, and, by convexity, for randomized ones as well.
\end{proof}
\noindent This concludes the proof of~\cref{lemma:nonuniformity:basic}.
\end{proof}
Given~\cref{lemma:nonuniformity:basic}, it would be tempting to try and establish~\cref{theo:nonuniformity} by (parallel) repetition, invoking~\cref{theo:sequential:repetition:mip}. While this would indeed lead to the desired sample and communication complexities (and only require $\dst_1<\dst_2$ instead of the more stringent $\dst_1<\dst_2^2$; though it would increase the number of rounds), it would unfortunately \emph{not} yield the claimed SZK property: intuitively, the idea is that a simulator which does not know $\totalvardist{\p}{\uniformOn{\ab}}$ \emph{exactly} in the completeness case (but only has a lower bound $\dst_2$ on it) can only simulate one instance of the protocol within total variation $\frac{1-\dst_2}{2}$.\footnote{To see why, note that the simulator, without knowledge of $\totalvardist{\p}{\uniformOn{\ab}}$, cannot ``know'' whether the probability to correctly guess $b$ should be the lower bound $\frac{1}{2}\Paren{1+\dst_2}$, or arbitrarily close to $1$.} After amplification over $T$ repetitions, the total variation between distributions of simulated and true transcripts could then grow roughly as $\Omega(\min(T(1-\dst_2),1))$.

\begin{proofof}{\cref{theo:nonuniformity}}
In order to obtain a meaningful Zero-Knowledge guarantee, we instead amplify the total variation distance in the completeness case (from $\dst_2$ to $1-\errprob$) \emph{before} running the basic protocol of~\cref{lemma:nonuniformity:basic}. To do so, we let \arthur run the mapping of the Polarization Lemma (\cref{lemma:polarization}), so that with 
\[
\Paren{\log(1/\errprob')}^{\bigO{1/\log(\dst_1^2/\dst_2)}}
\]
samples from $\p$ and $\uniformOn{\ab}$ (where $\errprob'$ is a parameter to be determined later, and we rely on the assumption $\dst_2^2 > \dst_1$ in order to apply \cref{lemma:polarization}), \arthur and \merlin can run~\cref{algo:nonuniformity:basic} on a distribution $\p'$ over a new domain of size $\ab'=\ab'(\ab,\dst_1,\dst_2,\errprob)$ (with new reference distribution $\p'^\ast$ fully determined by $\p$)\footnote{Note that $\p'^\ast$ will not, in general, be uniform on $[\ab']$: this is where we rely on the generalization of~\cref{lemma:nonuniformity:basic}.} such that (1) if $\totalvardist{\p}{\uniformOn{\ab}} \geq \dst_2$, then $\totalvardist{\p'}{\p'^\ast} \geq 1-2\errprob'$, and (2) if $\totalvardist{\p}{\uniformOn{\ab}} \leq \dst_1$, then $\totalvardist{\p'}{\p'^\ast} \leq 2\errprob'$.
Thereafter, we can apply $O(\log(1/\errprob))$ rounds of sequential repetition to reduce the probability in the soundness case with
\cref{theo:sequential:repetition:mip}.
To show that it is SZK (with deviation parameter $\errprob$), we give its simulator:
\begin{algorithm}[H]
    \begin{algorithmic}[1]
    
        \Require Sample access to $\p$
        \State $t \leftarrow \bigO{\log(1/\errprob)}$
        \ForAll{$i \gets 1$ to $t$}
        \LineComment{Store $b_i$ in memory to use at the end.}
        \State Chooses $b_i \samples \{0,1\}$
        \State Sample $x_i\sim \uniformOn{\ab}$ and $x'_i\sim \p$
        \EndFor
        \State Set the view to be $(b_1, \ldots, b_t, x_1, \ldots, x_t, x'_1, \ldots, x'_t, \hat{b}_1 = b_1, \ldots, \hat{b} = b_t)$
    \end{algorithmic}
    \caption{Simulator for \cref{algo:nonuniformity:basic} after sequential repetition}
\end{algorithm}
The transcript's total variation distance from the true transcript is exactly the probability that $\{\hat{b}_1\neq b_1\} \lor \cdots \lor \{\hat{b}_t\neq b_t\}$ in the latter, which is (by completeness) at most $\errprob' \cdot t = \errprob' \log (1 / \errprob)$ by a union bound. Setting $\errprob' = \errprob / \log(1 / \errprob)$ suffices. This establishes~\cref{theo:nonuniformity}.
\end{proofof}

\subsection{An unexpected consequence of~\cref{theo:nonuniformity}.}
It was shown in~\cite{Vadhan99} (Theorem~3.4.8) that \textsc{Entropy Difference} (ED) reduces to \textsc{Statistical Difference} (SD): the definition of the two computational problems assume that the input distributions are encoded as Boolean \emph{circuits} (so that sampling from $\p$ amounts to evaluating the circuit $\mathcal{C}_\p$ on a string $r$ of uniformly random input bits), and the reduction from ED to SD given in~\cite[Theorem~3.4.8]{Vadhan99} relies on this assumption. To the best of our knowledge, whether such a reduction is possible even in the weaker \emph{blackbox sampling} access model, was unknown.

A corollary of our results is that such a reduction given only sampling access to the distributions is impossible. To see why, recall that SD reduces to its complement: in more detail (see, \eg~\cite[Section~4.4]{Vadhan99}),
\begin{theorem}[{Reversing Statistical Difference~\cite[Corollary~4.4.1]{Vadhan99}}]
    \label{theo:reversing:statistical:distance}
There is a polynomial-time computable function that maps pairs of distributions $(\p,\q)$ (encoded as Boolean circuits) to pairs of distributions $(\p',\q')$ (encoded as Boolean circuits) such that
\begin{itemize}
    \item if $\totalvardist{\p}{\q} \geq \frac{2}{3}$, then $\totalvardist{\p'}{\q'} \leq \frac{1}{3}$;
    \item if $\totalvardist{\p}{\q} \leq \frac{1}{3}$, then $\totalvardist{\p'}{\q'} \geq \frac{2}{3}$.
\end{itemize}
\end{theorem}
The proof of this statement relies on combining three reductions: (1)~reduction from SD to ED, (2)~reduction from ED to its complement, and (3)~reduction from ED to SD. Of these, inspection of the proof shows that (1)~carries over to the blackbox sampling model, and so does (2)~(which is immediate). Moreover, all three steps are \emph{efficient}: for Boolean circuits with $m$ input gates and $n$ output gates (\ie inducing probability distributions over $\{0,1\}^n$), (1) and (2) only require a \emph{constant} number of circuit evaluations (or, equivalently, samples), while (3)~only requires $\poly(m,n)$ of them (a polylogarithmic dependence on the domain size $\ab=2^n$). 

From the above, it is apparent that \emph{if} one could obtain a version of (3), the reduction from ED to SD, in the blackbox sampling access model, using $t$ samples from $\p$ or $\q$, then the statistical difference reversal of~\cref{theo:reversing:statistical:distance} would be possible in the blackbox sampling access model as well, and require only $O(t)$ samples from $\p$ and $\q$ to sample from $\p'$ and $\q'$. But \emph{then}, one could use this reduction from SD to its complement on $(\p,\uniformOn{\ab})$ to reduce \emph{verifying uniformity} (a task known to require $\Omega(\sqrt{\ab})$ samples from \arthur) to \emph{verifying non-uniformity} (which, by~\cref{theo:nonuniformity}, can be done with $O(1)$ samples from \arthur)! This implies that any mapping reducing ED to SD in the blackbox sampling access model cannot be efficient, \ie must require $t=\Omega(\sqrt{\ab}) = \Omega(2^{n/2})$ samples from the input distributions:
\begin{corollary}
    \label{corollary:szk:reduction:hardness}
    Any reduction from \textsc{Entropy Difference} to \textsc{Statistical Difference} in the blackbox sampling access model (for probability distributions over $\{0,1\}^n$) requires $\Omega(2^{n/2})$ samples.
\end{corollary}

\newer{As an aside, the above result which shows the importance of the type of access allowed to the unknown distributions, leads to an interesting question regarding the \textsc{Statistical Difference} problem and its relation to the complexity class \textsf{SZK}. Indeed, the (blackbox) \emph{circuit access} is enough for this problem to be complete for \textsf{SZK}, while the above gives evidence that \emph{sample-only} access is not enough. But is the former entirely necessary?
\begin{openquestion}
    What is the minimal type of ``access'' to the distribution which preserves the completeness of \textsc{Statistical Difference} for  \textsf{SZK}? Is ``probability-revealing''~\cite{CanonneR14,OnakS18} access (whereby the algorithm gets pairs of the form $(x,\p(x))$, for $x$ randomly drawn from $\p$ or, generalizing, the uniform distribution) enough for this?
\end{openquestion}
}

\subsection{Some non-unexpected consequences of~\cref{theo:nonuniformity}.}
While the above was, for the sake of exposition, focusing on non-\emph{uniformity}, the algorithm and its proof immediately generalize to non-identity, non-closeness, and (not quite as immediately) non-independence. We now state these easy extensions.
\begin{corollary}[Verifying non-identity]
    \label{theo:nonidentity}
    There exists a one-round private-coin protocol $\Pi=\Pi(A,M)$ with the following guarantees: on input $\ab \geq 1$, constants $\errprob\in(0,1]$, $\dst_1,\dst_2\in[0,1]$ with $\dst_1 < \dst_2^2$, full knowledge by all parties of a known reference distribution $\p^\ast$, and sample access to an unknown probability distribution $\p\in\distribs{\ab}$,
    \begin{itemize}
        \item(Completeness) If $\totalvardist{\p}{\p^\ast} \geq \dst_2$, then $\probaOf{\Pi(A,M) = \accept } \geq 1-\errprob$;
        \item(Soundness) If $\totalvardist{\p}{\p^\ast} \leq \dst_1$, then, for all $M^\ast$, $\probaOf{\Pi(A,M^\ast) = \accept } \leq \errprob$;.
    \end{itemize}
    Moreover, $A$ takes $\Paren{\log(1/\errprob)}^{\bigO{1}} = \bigO{1}$ samples, and the protocol is (honest-verifier) Statistical Zero-Knowledge.
\end{corollary}
The proof is immediate by inspection, replacing $\uniformOn{\ab}$ by $\p^\ast$ in the proof of~\cref{theo:nonuniformity}. (One could also go the long and winding road to establish this result, leveraging the (now-standard) reduction from testing identity to an arbitrary fixed distribution $\q$ over a domain of size $\ab$ to that of testing uniformity over a domain of size $O(\ab)$ (see~\cite{DiakonikolasK16,Goldreich20}, or~\cite[Section~2.2.3]{Canonne22}). The only caveat is that the mapping underlying the aforementioned reduction only preserves total variation distances up to a constant factor, which then would slightly restrict the parameters in the completeness and soundness cases  to differ by at least a constant factor, \eg $\dst_1<\dst_2/2$.)\smallskip

\noindent One can also easily verify (!) that the above protocol extends to verifying farness between two unknown distributions \emph{in the one-prover case} (\ie when \merlin has full knowledge of both $\p,\q$). We emphasize that this is a setting we find less motivated, and differs from our study of farness in the two-prover case (\cref{sec:farness}):
\begin{corollary}[Verifying one-prover farness]
    \label{theo:noncloseness}
    There exists a one-round private-coin protocol $\Pi=\Pi(A,M)$ with the following guarantees: on input $\ab \geq 1$, , constants $\errprob\in(0,1]$, $\dst_1,\dst_2\in[0,1]$ with $\dst_1 < \dst_2^2$, and sample access to two unknown probability distribution $\p,\q\in\distribs{\ab}$ (both explicitly known by \merlin),
    \begin{itemize}
        \item(Completeness) If $\totalvardist{\p}{\q} \geq \dst_2$, then $\probaOf{\Pi(A,M) = \accept } \geq 1-\errprob$;
        \item(Soundness) If $\totalvardist{\p}{\q} \leq \dst_1$, then, for all $M^\ast$, $\probaOf{\Pi(A,M^\ast) = \accept } \leq \errprob$;.
    \end{itemize}
    Moreover, $A$ takes $\Paren{\log(1/\errprob)}^{\bigO{1}} = \bigO{1}$ samples, and the protocol is (honest-verifier) Statistical Zero-Knowledge.
\end{corollary}

Another nearly-immediate consequence of~\cref{theo:nonuniformity} is for \emph{non-independence verification}, whereby \merlin seeks to convince \arthur that the two variables $(X,Y)$ of a joint distribution $\p_{XY}$ are \emph{not} independent~--~an important task, \eg if one intends to fit a model to learn the statistical relation between two features of a dataset.
\begin{corollary}[Verifying non-independence]
  \label{lemma:nonindependence:basic}There exists a one-round private-coin
  protocol $(A, M)$, with the following guarantees: on input
  $\ab_X, \ab_Y \geq 1$, $\dst_1, \dst_2 \in [0, 1]$, $\dst_1 < \dst_2/3$ and sample access to an unknown
  probability distribution $\p_{X Y} \in \distribs{[\ab_X] \times [\ab_Y]}$,
  \begin{itemize}
    \item (Completeness) If $\totalvardist{\p_{X Y}}{ \q_X \otimes \q_Y} \geq
    \dst_2$ for every $\q_X \in \distribs{\ab_X}, \q_Y \in \distribs{\ab_Y}$, then \[
    \probaOf{\Pi (A, M) = \accept} \geq
    \frac{1}{2} + \frac{\dst_2}{2};
    \]
    
    \item (Soundness) If $\totalvardist{\p_{X Y}}{ \q_X \otimes \q_Y} \leq \dst_1$ for some $q_X \in \distribs{\ab_X}$ and $q_Y \in \distribs{\ab_Y}$, then,
    for all $M^{\ast}$, 
    \[
    \probaOf{\Pi (A, M^{\ast}) = \accept} \leq
    \frac{1}{2} + \frac{3 \dst_1}{2}.
    \]
  \end{itemize}
  Moreover, $A$ takes {\tmem{two}} samples, and the total communication is $\bigO{\log \ab_X + \log \ab_Y}$ bits.
\end{corollary}

\begin{proof}
  The proof follows from the fact that we can simulate $\p_X \otimes \p_Y$  given sample access to $\p_{X Y}$, simply by taking two independent samples $(x_1, y_1), (x_2, y_2) \sim \p_{X Y}$ and returning $(x_1, y_2)$. 

  \noindent For the completeness case, i.e., when $\p_{X Y}$ is far from independent, we note that 
  \[
  \totalvardist{\p_{XY}}{ \p_X \otimes \p_Y} \geq 
  \totalvardist{\p_{XY}}{\q_X \otimes \q_Y} \geq \dst_2 \,.
  \]
  \noindent For the soundness case, by \cite[Proposition 1]{BatuFFKRW01}, we have that $\totalvardist{\p_{XY}}{ \q_X \otimes \q_Y} \leq \dst_1$ implies
  \[
\totalvardist{\p_{X Y}}{\p_X \otimes \p_Y} \leq 3 \dst_1 
\]
  Therefore, checking farness of $\p_{XY}$ and $\p_X \otimes \p_Y$ suffice.
\end{proof}

\subsection{Application: Verifying farness from any efficiently learnable property} \label{sec:beyond:farness:uniformity}
Having obtained ultra-efficient, zero-knowledge proof systems for farness to uniformity and identity leads to a very natural question: \emph{is there more out there?} And indeed, were the answer negative, then this revisiting of interactive proof systems for distribution testing would have less appeal.

In this section, we provide a positive answer to this question, with a general result regarding farness verification of properties. At a high level, our result can be summarized as the verification counterpart of the \emph{testing-by-learning} framework (and its refinements, see, \eg~\cite{CanonneDGR18,AcharyaDK15}): ``if a property $\class$ can be efficiently \emph{learned}, then farness from it can be efficiently verified.'' (We also show that this statement is again specific to \emph{farness} from $\class$, and does not extend to \emph{closeness}, for which the $\Omega(\sqrt{\ab})$ sample complexity lower bound still applies.)

\begin{theorem}
    \label{theo:farness:property}
    Let $\class \subseteq \distribs{\ab}$ be a property of distributions which admits a semi-agnostic learner with sample complexity $q=q(\ab,\dst,\errprob)$ and semi-agnostic constant $C\geq 1$.\footnote{Recall that a semi-agnostic learner for a class $\class$ is an algorithm which, on input $\dst,\errprob$, takes \iid samples from an arbitrary distribution $\p$, and outputs a distribution $\hat{\p}$ such that, with probability at least $1-\errprob$, $\totalvardist{\p}{\hat{\p}} \leq C\cdot \min_{\q\in\class} \totalvardist{\p}{\q} + \dst$, where $C\geq 1$ is an absolute constant. If $C=1$, the learning algorithm is simply said to be \emph{agnostic}.} 
    Then, there exists a one-round private-coin protocol $\Pi=\Pi(A,M)$ with the following guarantees: on input $\ab \geq 1$, and constants $\errprob\in(0,1]$, $\dst_1,\dst_2\in[0,1]$ with \smash{$\dst_1 < \min(\frac{\dst_2}{2(C+2)},\frac{\dst_2^2}{4(C+1)})$}; and sample access to an unknown probability distribution $\p\in\distribs{\ab}$,
    \begin{itemize}
        \item(Completeness) If $\totalvardist{\p}{\class} \geq \dst_2$, then $\probaOf{\Pi(A,M) = \accept } \geq 1-\errprob$;
        \item(Soundness) If $\totalvardist{\p}{\class} \leq \dst_1$, then, for all $M^\ast$, $\probaOf{\Pi(A,M^\ast) = \accept } \leq \errprob$;.
    \end{itemize}
    Moreover, $A$ takes $q(\ab,\dst_1,\errprob/2)+\Paren{\log(1/\errprob)}^{\bigO{1/\log(\dst_1^2/\dst_2)}} = \bigO{q(\ab,\dst_1,\errprob/2)}$ samples, and the protocol is (honest-verifier) Statistical Zero-Knowledge.
\end{theorem}
As an example of such properties, we have that of \emph{monotone distributions} with ($q(\ab,\dst,\errprob) = O((\log(\ab/\errprob)/\dst^3)$), as well as, for instance, the classes of unimodal, $t$-modal, $t$-histograms, (discrete) log-concave, convex, concave, Monotone Hazard Rate (MHR), and $t$-piecewise degree-$d$ distributions (and mixtures thereof), which for $t,d=O(1)$ all have $\poly(1/\dst)$ or $\poly(\log(\ab),1/\dst)$-sample semi-agnostic learners~\cite{ChanDSS13,ChanDSS14}. This also applies to other structured classes, such as Poisson Binomial Distributions ($q(\ab,\dst,\errprob)=\tilde{O}(\log(1/\errprob)/\dst^2)$ by~\cite{DaskalakisDS15}) and Sums of Independent Integer-Valued random variables ($t$-SIIRVs) ($q(\ab,\dst,\errprob)=\poly(t,1/\dst)$ by~\cite{DaskalakisDOST13}).

We emphasize that we do not claim a full characterization of the properties whose farness can be efficiently verified. Our goals are much more modest: first, to show that there are \emph{many} such (natural) properties; and, second, to show how to use~\cref{theo:nonuniformity} (or, specifically,~\cref{theo:nonidentity}), \emph{in a blackbox way}, to verify farness from them.

\begin{proof}[Proof of~\cref{theo:farness:property}]
Fix any property \class satisfying the assumptions of the theorem, for the corresponding agnostic learning complexity $q$, and let $L$ be a semi-agnostic learning algorithm for \class achieving this sample complexity (and with semi-agnostic constant $C\geq 1$). The protocol then proceeds as follows:
\begin{enumerate}
    \item \arthur runs $L$ on samples from $\p$ with distance parameter $\dst_1$ and error probability $\frac{\errprob}{2}$, obtaining a hypothesis distribution $\hat{\p}$.
    \item \arthur then computes (without any samples~--~this is purely computational) $\totalvardist{\hat{\p}}{\class}$. If $\totalvardist{\hat{\p}}{\class} \geq \frac{1}{2}\dst_2$, it outputs \accept.
    \item Otherwise, \arthur sends the description of $\hat{\p}$ to \merlin,\footnote{Note that this distribution has a short description, as it was obtained from only $q$ samples, and so can be described with $O(q\log\ab)$ bits.} before running the farness protocol of~\cref{theo:nonidentity} on $\p$ (with reference $\hat{\p}$ and error probability $\frac{\errprob}{2}$) to verify that $\totalvardist{\p}{\hat{\p}} \geq \frac{\dst_2}{2}$ (versus $\totalvardist{\p}{\hat{\p}} \leq (C+1)\dst_1$). \arthur then only accepts if this protocol returns \accept.
\end{enumerate}
The sample complexity and SZK properties are immediate from the description of the protocol, and the guarantees of~\cref{theo:nonidentity} (note our assumptions on $\dst_1,\dst_2$ ensure that $(C+1)\dst_1 < (\dst_2/2)^2$). To establish correctness, we argue completeness and soundness separately. By a union bound, both the learning algorithm and the verification protocol simultaneously  satisfy their requirements with probability at least $1-\errprob$: we hereafter assume this holds.
\begin{description}
    \item[Completeness.] Assume $\totalvardist{\p}{\class} \geq \dst_2$. If $\totalvardist{\hat{\p}}{\class} > \frac{1}{2}\dst_2$, then \arthur accepts, and the protocol succeeds. If not, $\totalvardist{\hat{\p}}{\class} \leq \frac{1}{2}\dst_2$, and by the triangle inequality
    \[
        \totalvardist{\p}{\hat{\p}} \geq \dst_2  - \frac{1}{2}\dst_2 = \frac{\dst_2}{2}
    \]
    and so \merlin, in the farness protocol of~\cref{theo:nonidentity}, has a strategy leading to \accept.
    \item[Soundness.] Assume $\totalvardist{\p}{\class} \leq \dst_1$. Then, by the guarantees of the learning algorithm $L$, we have that
    \[
        \totalvardist{\p}{\hat{\p}} \leq C\cdot \totalvardist{\p}{\class}+\dst_1 \leq (C+1)\dst_1
    \]
    and in particular, by the triangle inequality,
    \[
        \totalvardist{\hat{\p}}{\class} \leq \totalvardist{\p}{\hat{\p}}+\totalvardist{\p}{\class} \leq (C+2)\dst_1 < \frac{1}{2}\dst_2\,,
    \]
    (the last inequality from our assumption on the parameter range) so \arthur does not output \accept in the second step.
    Moreover, since $\totalvardist{\p}{\hat{\p}} \leq (C+1)\dst_1$, \merlin, in the farness protocol of~\cref{theo:nonidentity}, has no strategy leading to \accept.
\end{description}
This concludes the proof.
\end{proof}

To conclude this section, we highlight that verifying \emph{closeness} (resp. \emph{membership}) to these properties, instead of \emph{farness} remains hard~--~\ie requires $\Omega(\sqrt{\ab})$ samples from \arthur (regardless of zero-knowledge). This follows by combining the hardness of verifying uniformity with the lower bound framework of~\cite[Section~6]{CanonneDGR18}, which carries over to verification: namely, any proof system for verifying (closeness or membership to) a property $\class$ which (1) contains the uniform distribution, and (2) admits an $o(\sqrt{\ab})$-sample semi-agnostic learning algorithm yields a proof system for verifying (closeness or membership to) uniformity.

\section{Verifying farness}
    \label{sec:farness}
In this section, we describe and analyze our two-prover protocols for verifying \emph{farness} of distributions: \merlin holds a distribution $\p$, \nimue a distribution $\q$, and they seek to convince \arthur that $\totalvardist{\p}{\q}$ is large. We begin by a very simple (but very much non-Zero-Knowledge) protocol in~\cref{sec:farness:leaking}, before building up to more satisfying (but similarly efficient) protocols in~\cref{ssec:farness:zk}: a perfect HIPOZK one for general distributions in~\cref{ssec:farness:zk}, and a perfect-ZK one (but for a very stringent parameter regime) in~\cref{ssec:farness:zk:disjoint}.




As a baseline, we recall that \emph{testing} farness (without a prover) is as hard as testing closeness, and thus requires $\smash{\bigOmega{\sqrt{\ab}/\dst^2 + \ab^{2/3}/\dst^{4/3}}}$ even to distinguish $\p=\q$ from $\totalvardist{\p}{\q} \geq \dst$. In the IP for distribution testing world, verifying \emph{closeness} also requires $\smash{\bigOmega{\sqrt{\ab}/\dst^2}}$ samples for \arthur, by the same simulation argument as for uniformity from~\cite{ChiesaG18}. As we will see, verifying \emph{farness} is again drastically easier~--~with \arthur only required to have a \emph{constant} (with respect to $\ab$) number of samples.

\subsection{Verifying farness is easy}
\label{sec:farness:leaking}
\begin{algorithm}[htbp]
    \begin{algorithmic}[1]
        \Require Sample access to $\p,\q$ (\arthur) over domain $\domain$, full knowledge of $\p$ (\merlin), full knowledge of $\q$ (\nimue)
        \LineComment{\textbf{\arthur}}
        \State Chooses $b \samples \{0,1\}$, 
        \State Samples $x_0\sim \p$ and $x_1\sim \q$
        \State Sends $x_b$ to \merlin
        \LineComment{\textbf{\merlin}}
        \State Upon receiving $x_b$, sends $\tau$ (real number $\p(x_b)$) to \arthur 
        \LineComment{\textbf{\arthur}}
        \State Upon receiving $\tau$, sends $x_b$ and $\tau$ to \nimue
        \LineComment{\textbf{\nimue}}
        \State Receives $\tau$, computes $\hat{b} \gets \indic{\tau < \q(x_b)}$.
        \State \Return \accept if $\hat{b}=b$.
    \end{algorithmic}
    \caption{Naive protocol for verifying farness}\label{algo:farness:leaking}
\end{algorithm}

\begin{theorem}
    \label{theo:farness}
    There exists a two-round private-coin protocol (\cref{algo:farness:leaking}) $\Pi=\Pi(A,M,N)$ with the following guarantees: on input $\ab \geq 1$, $\dst_1,\dst_2\in[0,1]$ with $\dst_1 < \dst_2$ and sample access to two unknown probability distributions $\p,\q\in\distribs{\ab}$ (where $A$ has oracle access to both, $M$ full knowledge of and access to $\p$ only, and $N$  full knowledge of and access to $\q$ only),
    \begin{itemize}
        \item(Completeness) If $\totalvardist{\p}{\q} \geq \dst_2$, then $\probaOf{\Pi(A,M,N) = \accept } \geq \frac{1 + \dst_2}{2}$;
        \item(Soundness) If $\totalvardist{\p}{\q} \leq \dst_1$, then, for all $M^\ast, N^\ast$, $\probaOf{\Pi(A,M^\ast,N^\ast) = \accept } \leq \frac{1 + \dst_1}{2}$;.
    \end{itemize}
    Moreover, $A$ takes $\bigO{1}$ samples.
\end{theorem}
\begin{proof}
    Since \nimue will have both "full" information (since \arthur will be relaying $\p(x_b)$) from both $\p$ and $\q$, this proof follows from single prover non-identity verification as in \cref{theo:nonidentity}.
\end{proof}

While the above result appears promising, it is unsatisfying in two ways: first, its communication complexity is unbounded, as sending the threshold $\tau$ (a real number) may require an arbitrarily large number of bits. (As we will see later in our ``better'' protocol (\cref{algo:farness:basic}), this issue can be relatively easily handled.) Second, and much more critical, it is very much \emph{not} zero-knowledge, and indeed blatantly fails to satisfy even the weaker notion of SK we introduced,~\cref{def:mip:distribution:opozk} in an inherent way: \merlin learns something about \nimue's distribution he could not have learned by himself, or even with the help of (sample-bounded) \arthur only.

\subsection{Verifying farness with Zero-Knowledge}
\label{ssec:farness:zk}

In this section, we establish the following result:
\begin{theorem}
    \label{theo:farness:general}
    There exists a $O(1)$-round private-coin protocol $\Pi=(A,M,N)$ with the following guarantees: on input $\ab \geq 1$, constants $\errprob\in(0,1]$, $\dst_1,\dst_2\in[0,1]$ with $\dst_1 < \dst_2$ and sample access to unknown probability distributions $\p,\q\in\distribs{\ab}$,
    \begin{itemize}
        \item(Completeness) If $\totalvardist{\p}{\q} \geq \dst_2$, then $\probaOf{\Pi(A,M,N) = \accept } \geq 1-\errprob$;
        \item(Soundness) If $\totalvardist{\p}{\q} \leq \dst_1$, then, for all $(M^\ast,N^\ast)$, $\probaOf{\Pi(A,M^\ast,N^\ast) = \accept } \leq \errprob$;.
    \end{itemize}
    Moreover, $A$ takes $O(1)$ sample from either $\p$ or $\q$, and the total communication is $\bigO{\log\ab}$ bits. In addition, this protocol achieves perfect HIPO Zero Knowledge.
\end{theorem}
The proof of this theorem will follow from the ``basic'' building block (\cref{lemma:farness:basic}) below, combined with the sequential repetition theorem (\cref{theo:sequential:repetition:mip}) (and the fact that $\errprob,\dst_1,\dst_2$ are constant).

\begin{algorithm}[htbp]
    \begin{algorithmic}[1]
        \Require Sample access to $\p,\q$ (\arthur) over domain $\domain$, full knowledge of $\p$ (\merlin), full knowledge of $\q$ (\nimue)
        \LineComment{\textbf{\arthur}}
        \State Chooses $b \samples \{0,1\}$, 
        \State Sample $x_0\sim \p$ and $x_1\sim \q$
        \State Send $x_b$ to both \merlin and \nimue  \label{algo:farness:basic:sendb} \Comment{$O(\log \ab)$ bits}
        \LineComment{\textbf{\merlin}}
        \State Upon receiving $x_b$, sends $\tau$ (an approximation of $\p(x_b)$) to \arthur \Comment{$O(\log\frac{\log\ab}{\dst_2-\dst_1})$ bits}
        \LineComment{\textbf{\nimue}}
        \State Upon receiving $x_b$, sends $\lambda$ (an approximation of $\q(x_b)$) to \arthur \Comment{$O(\log\frac{\log\ab}{\dst_2-\dst_1})$ bits}
        \LineComment{\textbf{\arthur}}
        \State Receives $\tau,\lambda$, computes $\hat{b} \gets \indic{\tau < \lambda}$.
        \State \Return \accept if $\hat{b}=b$.
    \end{algorithmic}
    \caption{Basic protocol for verifying farness}\label{algo:farness:basic}
\end{algorithm}

\begin{lemma}
    \label{lemma:farness:basic}
    There exists a one-round private-coin protocol $\Pi=(A,M,N)$,~\cref{algo:farness:basic}, with the following guarantees: on input $\ab \geq 1$, $\dst_1,\dst_2\in[0,1]$ with $\dst_1 < \dst_2$ and sample access to unknown probability distributions $\p,\q\in\distribs{\ab}$,
    \begin{itemize}
        \item(Completeness) If $\totalvardist{\p}{\q} \geq \dst_2$, then $\probaOf{\Pi(A,M,N) = \accept } \geq \frac{1}{2}+\frac{\dst_1+\dst_2}{4}$;
        \item(Soundness) If $\totalvardist{\p}{\q} \leq \dst_1$, then, for all $(M^\ast,N^\ast)$, $\probaOf{\Pi(A,M^\ast,N^\ast) = \accept } \leq \frac{1}{2}+\frac{\dst_1}{2}$;.
    \end{itemize}
    Moreover, $A$ takes \emph{one} sample from either $\p$ or $\q$, and the total communication is $\bigO{\log\frac{\ab}{\dst_2-\dst_1}}$ bits. In addition, this protocol achieves perfect HIPO Zero Knowledge.
\end{lemma}
\begin{proof}[Proof of~\cref{lemma:farness:basic}]
The lemma will directly follow from the next two claims, which, respectively, establish completeness and soundness of the protocol. Its sample complexity is immediate; the communication complexity will be addressed shortly.
\begin{claim}[Completeness]
    If $\totalvardist{\p}{\q} \geq \dst_2$, then there exists a strategy $M,N$ for \merlin and \nimue such that the protocol of~\cref{algo:farness:basic} returns \accept with probability at least $\frac{1+\frac{1}{2}(\dst_2+\dst_1)}{2}$.
\end{claim}
\begin{proof}
Let $\alpha \eqdef \frac{\dst_2-\dst_1}{4\dst_2}$, and denote by $\widetilde{S}_{\p,\q}$ the ``approximate Scheff\'e set'' corresponding to $\p$ and $\q$, defined as
\[
 \widetilde{S}_{\p,\q} \eqdef \setOfSuchThat{x\in \domain }{ \p(x) > \max\Paren{\Paren{1+\alpha\dst_2}\q(x), \frac{\alpha\dst_2}{\ab}}}
\]
and by $S_{\p,\q}$ the actual Scheff\'e set (\ie $S_{\p,\q}=S_{\p,\q}^0$). Then
\begin{align*}
\p(\widetilde{S}_{\p,\q}) - \q(\widetilde{S}_{\p,\q})
&= \p(S_{\p,\q}) - \q(S_{\p,\q}) - \Paren{\p(S_{\p,\q}\setminus \widetilde{S}_{\p,\q})-\q(S_{\p,\q}\setminus \widetilde{S}_{\p,\q})} \\
&\geq \dst_2 - \alpha\dst_2\q(S_{\p,\q}\setminus \widetilde{S}_{\p,\q}) - \frac{\alpha\dst_2}{\ab}\cdot \ab\\
&\geq (1-2\alpha)\dst_2 = \frac{1}{2}(\dst_2+\dst_1)\,.
\end{align*}
Considering the approximate Scheff\'e set instead of the ``true'' one allows \merlin and \nimue to be able to round the value $\tau,\lambda$ to send, as it suffices now to consider only $O(\log\frac{\log\ab}{\alpha\dst_2})=O(\log\frac{\log\ab}{\dst_2-\dst_1})$ possible values.\footnote{The encoding will work by first dividing $[0,1]$ into $O(\log \ab)$ buckets of a geometric series form: $\{[0, \frac{1}{\ab}], (\frac{1}{\ab}, \frac{2}{\ab}], \ldots, (\frac{2^{\lfloor \log \ab \rfloor}}{\ab}, 1]\}$ -- since we only need multiplicative approximation to encode $\Tilde{S}_{\p, \q}$. Then within each bucket, we will create uniform buckets of interval $\alpha \dst_2$. Encoding the location of the final bucket takes $\bigO{\log \frac{\log \ab}{\alpha \dst_2}}$ bits.} This enables the stated communication complexity.
The probability of acceptance is then
\begin{align*}
        \probaOf{\hat{b}=b} 
        &= \probaCond{\tau < \lambda}{b=0}\probaOf{b=0}+ \probaCond{\tau \geq \lambda}{b=1}\probaOf{b=1} \\
        &= \frac{1}{2}\Paren{ \probaOf{x_0 \in \widetilde{S}_{\p,\q} } + \probaOf{x_1 \notin \widetilde{S}_{\p,\q} } }\\
        &= \frac{1}{2}\Paren{ \p(\widetilde{S}_{\p,\q}) + 1 - \q(\widetilde{S}_{\p,\q}) }\\
        &\geq \frac{1}{2}\Paren{1+\frac{1}{2}(\dst_2+\dst_1) }\,,
    \end{align*}
    as claimed.
\end{proof}

\begin{claim}[Soundness]
    If $\totalvardist{\p}{\q} \leq \dst_1$, then there is no strategy $M,N$ for \merlin and \nimue such that the protocol of~\cref{algo:farness:basic} returns \accept with probability greater than $\frac{1+\dst_1}{2}$.
\end{claim}
\begin{proof}
    This again follows from the characterization of total variation distance as best possible advantage for distinguishing two fixed distributions given a single observation. Namely, one can consider a ``friendly but forgetful'' \arthur which does not remember $b$ after~\cref{algo:farness:basic:sendb}, but gets the full knowledge of $\p,\q$ from \merlin and \nimue to help him figure it out: then, any strategy $M,N$ yields a strategy for \arthur to guess $b$ from a single sample from a distribution promised to be either $\p$ or $\q$. The best probability of success for this augmented \arthur is $\frac{1}{2}(1+\totalvardist{\p}{\q})$.
\end{proof}

\begin{claim}[Perfect HIPOZK]
There exists a simulator $S^{\p, \q}(\p, t)$, when given sample access access to $\p, \q$ (sample complexity same as $\Pi$), and $\p_i$'s full description, can simulate $M$'s view of $\Pi$ in the completeness case
    \[
        \totalvardist{S^{\p, \q}(\p,t)}{\dotprod{A^{\p,\q}}{M(\p), N(\q))}_{M}(t)} = 0,
    \]
\end{claim}
\begin{algorithm}[H]
    \begin{algorithmic}[1]
        \Require Sample access to $\p,\q$ and $\p$'s full description.
        \State Chooses $b \samples \{0,1\}$
        \State Sample $x_0\sim \p$ and $x_1\sim \q$
        \State Set the view to be $(x_b, \p(x_b))$
    \end{algorithmic}
    \caption{\label{algo:farness:basic:simulator}Simulator for \cref{algo:farness:basic}}
\end{algorithm}
\begin{proof}
    It is not hard to see that \cref{algo:farness:basic:simulator} exactly simulates Merlin's view.
\end{proof}

\noindent This concludes the proof of~\cref{lemma:farness:basic}.
\end{proof}

Note that the above protocol, while (perfect) HIPOZK, is \emph{not} SZK. We leave as an open problem whether one can achieve SZK for farness verification of general distributions:
\begin{openquestion}
    \label{openproblem:szk:farness}
    Can one obtain similar guarantees as in~\cref{theo:farness:general} (farness verification for general distributions), but with \emph{Statistical ZK}? As a first step, can we obtain an SZK protocol for the specific case of \emph{flat} distributions?
\end{openquestion}
\subsection{Is Perfect Zero-Knowledge a dream?}
\label{ssec:farness:zk:disjoint}
As a counterpoint to the above, we now provide a protocol for farness (in a very specific, yet non-trivial parameter regime) which {does} provide \emph{perfect} ZK, showing that this requirement is indeed achievable.
\begin{theorem}
    \label{lemma:farness:zk:perfectzk}
    There exists a one-round private-coin protocol $\Pi=\Pi(A,M,N)$ with the following guarantees: on input $\ab \geq 1$, $\dst\in[0,1)$ and sample access to two unknown probability distributions $\p,\q\in\distribs{\ab}$ (where $A$ has oracle access to both, $M$ full knowledge of and access to $\p$ only, and $N$  full knowledge of and access to $\q$ only),
    \begin{itemize}
        \item(Completeness) If $\totalvardist{\p}{\q} = 1$, then $\probaOf{\Pi(A,M,N) = \accept } = 1$;
        \item(Soundness) If $\totalvardist{\p}{\q} \leq \dst$, then, for all $M^\ast, N^\ast$, $\probaOf{\Pi(A,M^\ast,N^\ast) = \accept } \leq \frac{1}{2} + \frac{\dst}{2}$;.
    \end{itemize}
    Moreover, $A$ takes \emph{one} sample, and the protocol is (honest-verifier) \emph{Perfect} Zero-Knowledge.
\end{theorem}

\begin{algorithm}[htbp]
    \begin{algorithmic}[1]
        \Require Sample access to $\p,\q$ (\arthur) over domain $\domain$, full knowledge of $\p$ (\merlin), full knowledge of $\q$ (\nimue), distance parameter $\dst$.
        \LineComment{\textbf{\arthur}}
        \State Chooses $b \samples \{0,1\}$
        \State Sample $x_0\sim \p$ and $x_1\sim \q$
        \State Send $x_b$ to both \merlin and \nimue \Comment{$O(\log \ab)$ bits}
        \LineComment{\textbf{\merlin}}
        \State Upon receiving $x_b$, sends $\hat{b}_1$ (a guess for $b$) to \arthur
        \LineComment{\textbf{\nimue}}
        \State Upon receiving $x_b$, sends $\hat{b}_2$ (a guess for $b$) to \arthur
        \LineComment{\textbf{\arthur}}
        \State \Return \accept if $b=\hat{b}_1=\hat{b}_2$.
    \end{algorithmic}
    \caption{Perfect Zero Knowledge protocol for verifying disjointness}\label{algo:farness:total}
\end{algorithm}
\begin{proof}
    We provide the completeness, soundness, and zero knowledge analysis below; the protocol is given in~\cref{algo:farness:total}.
    \paragraph{Completeness.} \merlin and \nimue's strategy is to send the indicator of whether the received sample belong to the support of their distribution: that is, $\hat{b}_1 = \indicSet{\supp{\p}}(x_b)$ and $\hat{b}_2 = \indicSet{\supp{\q}}(x_b)$. If $\totalvardist{\p}{\q}=1$, then $\supp{\p}\cap \supp{\q} = \emptyset$ and as a result $\probaOf{b=\hat{b}_1=\hat{b}_2} = 1$, as desired.

    \paragraph{Soundness.} Here again soundness directly follows from the hypothesis characterization of total variation distance, see \cref{claim:basicsoundness}.

    \paragraph{Perfect Zero-Knowledge.} There is an obvious simulator for \arthur's view of the interaction in the completeness case:
\begin{algorithm}[H]
    \begin{algorithmic}[1]
        \Require Sample access to $\p,\q$
        \State Chooses $b \samples \{0,1\}$
        \State Sample $x_0\sim \p$ and $x_1\sim \q$
        \State Set the view to be $(b,x_0,x_1, x_b, \hat{b}_1=b, \hat{b}_2=b)$
    \end{algorithmic}
    \caption{Simulator for \cref{algo:farness:total}}
\end{algorithm}
The transcript (over the choice of $b, x_0,x_1$) has then exactly the same distribution as \arthur's view of a honest interaction with \merlin and \nimue in the completeness case.
\end{proof}


\subsection{Application: Verifying non-clusterability} \label{sec:farness:clusterability}
To conclude this section, we describe how to use a farness verification protocol (such as the one of~\cref{theo:farness:general}), as a blackbox, to verify \emph{non-clusterability}. Recall that a collection of $\ell$ distributions $\p_1,\dots,\p_\ell \in \distribs{\ab}$ is said to be \emph{$(t,\alpha)$}-clusterable if there exists a partition of $[\ell]$ into $t$ lists $L_1,\dots,L_t$ such that, for every $r\in[t]$ and every $i,j\in L_r$, $\totalvardist{\p_i}{\p_j} \leq \alpha$~\cite[Definition~7.1]{LeviRR13}

For $\alpha=O(1/\sqrt{\ab})$, Levi, Ron, and Rubinfeld provided in~\cite{LeviRR13} an algorithm for testing $(t,\alpha)$-clusterability in the \emph{query model} (where the algorithm can choose, at each step, which of the $\ell$ distributions to sample) which, for constant $\dst$, has sample complexity $\tilde{O}(\ell \cdot \ab^{2/3})$. Moreover, any testing algorithm must have sample complexity $\Omega(\ab^{2/3})$, even for $\alpha=0$.\footnote{Moreover, as before, the by-now usual simulation argument carries over to show that $\Omega(\sqrt{\ab})$ samples are necessary from \arthur to verify $(t,\alpha)$-clusterability, even for $t=2$ and $\alpha=O(1/\sqrt{\ab})$.)}

We show how to adapt the algorithm of~\cite{LeviRR13} to the multiprover setting, to verify farness from $(t,\alpha)$-clusterability, with no restriction on the range of $\alpha$.
\begin{theorem}
    \label{theo:clusterability}
    There exists a $t$-round private-coin protocol $\Pi=\Pi(A,M_1,\dots,M_\ell)$ with the following guarantees: on input $\ab \geq 1$, $1\leq t\leq \ell$,  constants $\dst\in[0,1)$, $\errprob\in(0,1]$, $0\leq \alpha < \frac{1}{2}\dst$ and sample access to $\ell$ unknown probability distributions $\p_1,\dots, \p_\ell\in\distribs{\ab}$ (where $A$ has oracle access to all, and each $M_i$ full knowledge of and access to $\p_i$ only),
    \begin{itemize}
        \item(Completeness) If $\p_1,\dots, \p_\ell$ are $\dst$-far from $(t,\alpha)$-clusterable, then $\probaOf{\Pi(A,M_1,\dots,M_\ell) = \accept } \geq 1-\errprob$;
        \item(Soundness) If $\p_1,\dots, \p_\ell$ are $(t,\alpha)$-clusterable, then, for all prover strategies $M_1^\ast, \dots,M_\ell^\ast$, we have $\probaOf{\Pi(A,M_1^\ast, \dots,M_\ell^\ast) = \accept } \leq \errprob$;.
    \end{itemize}
    Moreover, $A$ takes $\tildeO{\newer{t^2}}$ samples, and the protocol is (honest-verifier) Statistical Zero-Knowledge.
\end{theorem}

\begin{proof}[Proof sketch]
The algorithm is a relatively straightforward adaptation of~\cite[Algorithm~5]{LeviRR13}, using our farness verification protocol as a building block instead of a tolerant testing algorithm:
\begin{algorithm}[htbp]
    \begin{algorithmic}[1]
        \State \arthur picks the first representative, $r_1$, u.a.r. from $[\ell]$;
        \ForAll{$1\leq i\leq t$}
            \State \arthur selects a set $S$ of $O(\log(t)/\dst)$ indices u.a.r. from $[\ell]$
            \ForAll{$j\in S$}
                \ForAll{$1\leq i'\leq i$}
                    \State Run the farness verification protocol (\cref{theo:farness:general}) with $M_{r_{i'}}$ and $M_j$ (on $\p_{r_{i'}}$ and $\p_j$) with $\dst_1 = \alpha$, $\dst_2 = \frac{\dst}{2}$, and error probability $\frac{\errprob}{(t+1)^2 |S|}$
                \EndFor
                \If{there is some $j\in S$ for which all $i$ verification protocols returned \accept}
                    \State Set $r_{i+1} = j$ \Comment{$\p_j$ is far from all current clusters}
                    \State Continue to the $(i+1)$-th iteration of the outer loop
                \Else 
                    \State\Return \reject \Comment{no distribution requiring opening a new cluster was found}
                \EndIf
            \EndFor
        \EndFor
    \State\Return \accept\label{algo:clusterability:accept}
    \end{algorithmic}
    \caption{Protocol for verifying non-clusterability}
\end{algorithm}
The claimed round and sample complexities (for constant $\alpha,\errprob$) are immediate from the algorithm and \cref{theo:farness:general}. Turning to the correctness: by a union bound, all runs of the farness verification protocol will be successful overall with probability at least $\errprob$: assume this event holds.

\begin{description}
    \item[Completeness.] Suppose $\p_1,\dots, \p_\ell$ are $\dst$-far from $(t,\alpha)$-clusterable. Then, the analysis from~\cite[Algorithm~5]{LeviRR13} (their Theorem~7.2) establishes that as long as $i \leq t$ cluster representatives have been found, the number of distributions among $\p_1,\dots,\p_\ell$ which are $\dst/2$ far from all current representatives is at least $\dst/(2\ell)$, and so, from the setting of $|S|$ and by a union bound, with high probability at least one will be included, at each iteration of the outer loop, in the randomly selected set $S$~--~and will then be verified to be far when running the corresponding farness verification protocol, becoming the next representative. Thus, Line~\ref{algo:clusterability:accept} will be reached at the end, and the protocol will return \accept.
    \item[Soundness.] Assume now that $\p_1,\dots, \p_\ell$ are $(t,\alpha)$-clusterable. Since the protocol only returns \accept if at least $t+1$ cluster centers are found, the only way this outcome is reached is if at least $t+1$ farness verification protocols return \accept (even though by assumption there is no pairwise $\alpha$-far set of $t+1$ distributions). But this cannot happen, as this would violate the soundness of this protocol.
\end{description}
This concludes the proof.
\end{proof}

\begin{openquestion}
    \label{openproblem:collections}
    Which other properties considered in the collection of distributions setting are amenable to ultra-efficient verification protocols~--~and which general statements can be made on the connection between the two models?
\end{openquestion}

\section{Repetition theorems}
    \label{sec:repetition}
\begin{theorem}[Sequential repetition for MIP distribution verification]
\label{theo:sequential:repetition:mip}
    Suppose an MIP proof system $\Pi(M_1(\p_1),\dots,M_\ell(\p_\ell),A^{\p_1,\dots,\p_\ell})$ for distribution verification of properties $\cP_0,\cP_1$ exists with completeness parameter $1/2 + \Delta_c$ and soundness parameter $1/2 + \Delta_s$, where $\Delta_c > \Delta_s$. Then, for any $t \in \N$, there exists an MIP proof system $\Pi'(M_1(\p_1),\dots,M_\ell(\p_\ell),A^{\p_1,\dots,\p_\ell})$ for verification of properties $\cP_0,\cP_1$ with completeness parameter $1-2^{-\Omega(t)}$ and soundness parameter $2^{-\Omega(t)}$, where $\Pi'$ is obtained by sequentially repeating the original MIP protocol $\bigO{t/(\Delta_c-\Delta_s)^2}$ times.
\end{theorem}
\begin{proof}[Proof Sketch]
    The proof follows as for sequential repetition of IP protocols in the functional setting (see, in particular, Theorem~1.4 in \cite{Prashant21Lec03}). Sequential repetition with $m$ repetitions works as follows: Run the protocol $\Pi$ sequentially $m$ times, and let $\rho$ be the fraction of times the protocol $\Pi$ accepts. Then the overall protocol $\Pi'$ accepts if and only if $\rho \geq (1+ \Delta_c - \Delta_s)/2 $. 

    \noindent \textit{Completeness:} If the provers treat each of the $m$ repetitions of $\Pi$ independently, then each repetition outputs \accept with probability at least $1/2+\Delta_c$ independently of the other rounds. Thus, by Chernoff bound, it follows that $\Pi'$ accepts with probability $1-2^{-\Omega(m)} = 1 - 2^{-\Omega(t)}$.

    \noindent \textit{Soundness:}
    As in the classical functional setting, the soundness guarantee follows from the fact that if the soundness error of $\Pi'$ (i.e. $m$ sequential repetitions of $\Pi$) does not decay as desired (i.e. does not decay as if the provers behaved independently across the $m$ repetitions), then there exists a cheating prover strategy that will cause the original protocol $\Pi$ to (wrongly) output \accept on an instance from $\cP_0$ with probability $> 1/2+ \Delta_s$ (see Claim~1.2 and its proof in \cite{Prashant21Lec03}). This would contradict the soundness guarantee of $\Pi$.

    \end{proof}

\begin{openquestion}[Parallel repetition for MIP distribution verification] 
We conjecture that the thresholding protocol as in the sequential repetition above leads to an exponential decay in soundness parameter even with parallel repetition in the single-prover scenario following arguments in \cite{Goldreich25}. 
\end{openquestion}

\printbibliography

@inproceedings{Tell16,
  author       = {Roei Tell},
  title        = {On Being Far from Far and on Dual Problems in Property Testing: [Extended
                  Abstract]},
  booktitle    = {{ITCS}},
  pages        = {103--110},
  publisher    = {{ACM}},
  year         = {2016}
}

@inproceedings{ValiantV11,
  author       = {Gregory Valiant and
                  Paul Valiant},
  title        = {Estimating the unseen: an n/log(n)-sample estimator for entropy and
                  support size, shown optimal via new CLTs},
  booktitle    = {{STOC}},
  pages        = {685--694},
  publisher    = {{ACM}},
  year         = {2011}
}

@article{ErgunKR04,
  author       = {Funda Erg{\"{u}}n and
                  Ravi Kumar and
                  Ronitt Rubinfeld},
  title        = {Fast approximate probabilistically checkable proofs},
  journal      = {Inf. Comput.},
  volume       = {189},
  number       = {2},
  pages        = {135--159},
  year         = {2004},
  doi          = {10.1016/J.IC.2003.09.005},
}

@inproceedings{BenOrGKW88,
  author       = {Michael Ben{-}Or and
                  Shafi Goldwasser and
                  Joe Kilian and
                  Avi Wigderson},
  title        = {Multi-Prover Interactive Proofs: How to Remove Intractability Assumptions},
  booktitle    = {{STOC}},
  pages        = {113--131},
  publisher    = {{ACM}},
  year         = {1988}
}

@book{Goldreich98,
  author       = {Oded Goldreich},
  title        = {Modern Cryptography, Probabilistic Proofs and Pseudorandomness},
  series       = {Algorithms and Combinatorics},
  volume       = {17},
  publisher    = {Springer},
  year         = {1998}
}

@article{goldreich1998property,
  author       = {Oded Goldreich and
                  Shafi Goldwasser and
                  Dana Ron},
  title        = {Property Testing and its Connection to Learning and Approximation},
  journal      = {J. {ACM}},
  volume       = {45},
  number       = {4},
  pages        = {653--750},
  year         = {1998},
  doi          = {10.1145/285055.285060},
}

@article{RubinfeldS96,
  author       = {Ronitt Rubinfeld and
                  Madhu Sudan},
  title        = {Robust Characterizations of Polynomials with Applications to Program
                  Testing},
  journal      = {{SIAM} J. Comput.},
  volume       = {25},
  number       = {2},
  pages        = {252--271},
  year         = {1996},
  doi          = {10.1137/S0097539793255151}
}

@article{Fischer01,
  author       = {Eldar Fischer},
  title        = {The Art of Uninformed Decisions: A Primer to Property Testing},
  journal      = {Bull. {EATCS}},
  volume       = {75},
  pages        = {97},
  year         = {2001},
    doi          = {10.1142/9789812562494_0014}
}

@incollection{fischer2004art,
  title={The art of uninformed decisions: A primer to property testing},
  author={Fischer, Eldar},
  booktitle={Current Trends in Theoretical Computer Science: The Challenge of the New Century Vol 1: Algorithms and Complexity Vol 2: Formal Models and Semantics},
  pages={229--263},
  year={2004},
  publisher={World Scientific}
}

@article{Ron08,
  author       = {Dana Ron},
  title        = {Property Testing: {A} Learning Theory Perspective},
  journal      = {Found. Trends Mach. Learn.},
  volume       = {1},
  number       = {3},
  pages        = {307--402},
  year         = {2008},
  url          = {https://doi.org/10.1561/2200000004},
  doi          = {10.1561/2200000004},
}

@book{Goldreich10,
  editor       = {Oded Goldreich},
  title        = {Property Testing - Current Research and Surveys},
  series       = {Lecture Notes in Computer Science},
  volume       = {6390},
  publisher    = {Springer},
  year         = {2010},
  url          = {https://doi.org/10.1007/978-3-642-16367-8},
  doi          = {10.1007/978-3-642-16367-8},
  isbn         = {978-3-642-16366-1}
}

@book{Goldreich17,
  author       = {Oded Goldreich},
  title        = {Introduction to Property Testing},
  publisher    = {Cambridge University Press},
  year         = {2017},
  url          = {http://www.cambridge.org/us/catalogue/catalogue.asp?isbn=9781107194052},
  doi          = {10.1017/9781108135252}
}

@inproceedings{batu2000testing,
  title={Testing that distributions are close},
  author={Batu, Tugkan and Fortnow, Lance and Rubinfeld, Ronitt and Smith, Warren D and White, Patrick},
  booktitle={Proceedings 41st Annual Symposium on Foundations of Computer Science},
  pages={259--269},
  year={2000},
  organization={IEEE}
}

@article{BatuFRS10,
  author       = {Tugkan Batu and
                  Lance Fortnow and
                  Ronitt Rubinfeld and
                  Warren D. Smith and
                  Patrick White},
  title        = {Testing Closeness of Discrete Distributions},
  journal      = {CoRR},
  volume       = {abs/1009.5397},
  year         = {2010}
}

@article{Canonne22,
  author       = {Cl{\'{e}}ment L. Canonne},
  title        = {Topics and Techniques in Distribution Testing: {A} Biased but Representative
                  Sample},
  journal      = {Found. Trends Commun. Inf. Theory},
  volume       = {19},
  number       = {6},
  pages        = {1032--1198},
  year         = {2022}
}

@inproceedings{RothblumVW13,
  author       = {Guy N. Rothblum and
                  Salil P. Vadhan and
                  Avi Wigderson},
  editor       = {Dan Boneh and
                  Tim Roughgarden and
                  Joan Feigenbaum},
  title        = {Interactive proofs of proximity: delegating computation in sublinear
                  time},
  booktitle    = {Symposium on Theory of Computing Conference, STOC'13, Palo Alto, CA,
                  USA, June 1-4, 2013},
  pages        = {793--802},
  publisher    = {{ACM}},
  year         = {2013},
  doi          = {10.1145/2488608.2488709},
}

@inproceedings{ChiesaG18,
  author       = {Alessandro Chiesa and
                  Tom Gur},
  editor       = {Anna R. Karlin},
  title        = {Proofs of Proximity for Distribution Testing},
  booktitle    = {9th Innovations in Theoretical Computer Science Conference, {ITCS}
                  2018, January 11-14, 2018, Cambridge, MA, {USA}},
  series       = {LIPIcs},
  volume       = {94},
  pages        = {53:1--53:14},
  year         = {2018},
  url          = {https://doi.org/10.4230/LIPIcs.ITCS.2018.53},
  doi          = {10.4230/LIPICS.ITCS.2018.53},
}

@inproceedings{GoldwasserRSY21,
  author       = {Shafi Goldwasser and
                  Guy N. Rothblum and
                  Jonathan Shafer and
                  Amir Yehudayoff},
  editor       = {James R. Lee},
  title        = {Interactive Proofs for Verifying Machine Learning},
  booktitle    = {12th Innovations in Theoretical Computer Science Conference, {ITCS}
                  2021, January 6-8, 2021, Virtual Conference},
  series       = {LIPIcs},
  volume       = {185},
  pages        = {41:1--41:19},
  publisher    = {Schloss Dagstuhl - Leibniz-Zentrum f{\"{u}}r Informatik},
  year         = {2021},
  doi          = {10.4230/LIPICS.ITCS.2021.41}
}

@inproceedings{BermanRV18,
  author       = {Itay Berman and
                  Ron D. Rothblum and
                  Vinod Vaikuntanathan},
  editor       = {Anna R. Karlin},
  title        = {Zero-Knowledge Proofs of Proximity},
  booktitle    = {9th Innovations in Theoretical Computer Science Conference, {ITCS}
                  2018, January 11-14, 2018, Cambridge, MA, {USA}},
  series       = {LIPIcs},
  volume       = {94},
  pages        = {19:1--19:20},
  publisher    = {Schloss Dagstuhl - Leibniz-Zentrum f{\"{u}}r Informatik},
  year         = {2018},
  doi          = {10.4230/LIPICS.ITCS.2018.19}
}

@inproceedings{HermanR22,
  author       = {Tal Herman and
                  Guy N. Rothblum},
  editor       = {Stefano Leonardi and
                  Anupam Gupta},
  title        = {Verifying the unseen: interactive proofs for label-invariant distribution
                  properties},
  booktitle    = {{STOC} '22: 54th Annual {ACM} {SIGACT} Symposium on Theory of Computing,
                  Rome, Italy, June 20 - 24, 2022},
  pages        = {1208--1219},
  publisher    = {{ACM}},
  year         = {2022},
  doi          = {10.1145/3519935.3519987}
}

@inproceedings{HermanR23,
  author       = {Tal Herman and
                  Guy N. Rothblum},
  title        = {Doubley-Efficient Interactive Proofs for Distribution Properties},
  booktitle    = {64th {IEEE} Annual Symposium on Foundations of Computer Science, {FOCS}
                  2023, Santa Cruz, CA, USA, November 6-9, 2023},
  pages        = {743--751},
  publisher    = {{IEEE}},
  year         = {2023},
  doi          = {10.1109/FOCS57990.2023.00049}
}

@inproceedings{HermanR24,
  author       = {Tal Herman and
                  Guy N. Rothblum},
  title        = {Interactive Proofs for General Distribution Properties},
  booktitle    = {65th {IEEE} Annual Symposium on Foundations of Computer Science, {FOCS}
                  2024, Chicago, IL, USA, October 27-30, 2024},
  pages        = {528--538},
  publisher    = {{IEEE}},
  year         = {2024},
  doi          = {10.1109/FOCS61266.2024.00041}
}

@inproceedings{HermanR25,
  author       = {Tal Herman and
                  Guy N. Rothblum},
  title        = {How to Verify Any (Reasonable) Distribution Property: Computationally
                  Sound Argument Systems for Distributions},
  booktitle    = {The Thirteenth International Conference on Learning Representations,
                  {ICLR} 2025, Singapore, April 24-28, 2025},
  publisher    = {OpenReview.net},
  year         = {2025}
}

@article{GoldreichRothblumG25,
  author       = {Oded Goldreich and
                  Guy N. Rothblum},
  title        = {Location-Invariant Properties of Functions versus Properties of Distributions: United in Testing but Separated in Verification},
  journal      = {Electron. Colloquium Comput. Complex.},
  volume       = {{TR25-105}},
  year         = {2025},
}

@article{GoldreichHermanRothblum25,
  author       = {Oded Goldreich and
                  Tal Herman and
                  Guy N. Rothblum},
  title        = {Interactive proof systems for {FARNESS}},
  journal      = {Electron. Colloquium Comput. Complex.},
  volume       = {{TR25-201}},
  year         = {2025},
  url          = {https://eccc.weizmann.ac.il/report/2025/201},
  eprinttype    = {ECCC},
}

@article{GoldreichRon99,
  author       = {Oded Goldreich and
                  Dana Ron},
  title        = {A Sublinear Bipartiteness Tester for Bounded Degree Graphs},
  journal      = {Comb.},
  volume       = {19},
  number       = {3},
  pages        = {335--373},
  year         = {1999}
}

@article{Paninski08,
  author       = {Liam Paninski},
  title        = {A Coincidence-Based Test for Uniformity Given Very Sparsely Sampled
                  Discrete Data},
  journal      = {{IEEE} Trans. Inf. Theory},
  volume       = {54},
  number       = {10},
  pages        = {4750--4755},
  year         = {2008},
  url          = {https://doi.org/10.1109/TIT.2008.928987},
  doi          = {10.1109/TIT.2008.928987},
}

@inproceedings{BatuFRSW00,
  author       = {Tugkan Batu and
                  Lance Fortnow and
                  Ronitt Rubinfeld and
                  Warren D. Smith and
                  Patrick White},
  title        = {Testing that distributions are close},
  booktitle    = {41st Annual Symposium on Foundations of Computer Science, {FOCS} 2000,
                  Redondo Beach, California, USA, November 12-14, 2000},
  pages        = {259--269},
  publisher    = {{IEEE} Computer Society},
  year         = {2000},
  url          = {https://doi.org/10.1109/SFCS.2000.892113},
  doi          = {10.1109/SFCS.2000.892113}
}

@article{Valiant11,
  author       = {Paul Valiant},
  title        = {Testing Symmetric Properties of Distributions},
  journal      = {{SIAM} J. Comput.},
  volume       = {40},
  number       = {6},
  pages        = {1927--1968},
  year         = {2011},
  url          = {https://doi.org/10.1137/080734066},
  doi          = {10.1137/080734066}
}

@article{ChanDVV13,
  author       = {Siu On Chan and
                  Ilias Diakonikolas and
                  Gregory Valiant and
                  Paul Valiant},
  title        = {Optimal Algorithms for Testing Closeness of Discrete Distributions},
  journal      = {CoRR},
  volume       = {abs/1308.3946},
  year         = {2013},
  url          = {http://arxiv.org/abs/1308.3946},
  eprinttype    = {arXiv},
  eprint       = {1308.3946}
}

@article{BatuFRSW13,
  author       = {Tugkan Batu and
                  Lance Fortnow and
                  Ronitt Rubinfeld and
                  Warren D. Smith and
                  Patrick White},
  title        = {Testing Closeness of Discrete Distributions},
  journal      = {J. {ACM}},
  volume       = {60},
  number       = {1},
  pages        = {4:1--4:25},
  year         = {2013},
  url          = {https://doi.org/10.1145/2432622.2432626},
  doi          = {10.1145/2432622.2432626}
}

@inproceedings{BatuKR04,
  author       = {Tugkan Batu and
                  Ravi Kumar and
                  Ronitt Rubinfeld},
  editor       = {L{\'{a}}szl{\'{o}} Babai},
  title        = {Sublinear algorithms for testing monotone and unimodal distributions},
  booktitle    = {Proceedings of the 36th Annual {ACM} Symposium on Theory of Computing,
                  Chicago, IL, USA, June 13-16, 2004},
  pages        = {381--390},
  publisher    = {{ACM}},
  year         = {2004},
  url          = {https://doi.org/10.1145/1007352.1007414},
  doi          = {10.1145/1007352.1007414}
}

@article{CanonneDGR18,
  author       = {Cl{\'{e}}ment L. Canonne and
                  Ilias Diakonikolas and
                  Themis Gouleakis and
                  Ronitt Rubinfeld},
  title        = {Testing Shape Restrictions of Discrete Distributions},
  journal      = {Theory Comput. Syst.},
  volume       = {62},
  number       = {1},
  pages        = {4--62},
  year         = {2018}
}

@article{GuhaMV09,
  author       = {Sudipto Guha and
                  Andrew McGregor and
                  Suresh Venkatasubramanian},
  title        = {Sublinear estimation of entropy and information distances},
  journal      = {{ACM} Trans. Algorithms},
  volume       = {5},
  number       = {4},
  pages        = {35:1--35:16},
  year         = {2009},
  url          = {https://doi.org/10.1145/1597036.1597038},
  doi          = {10.1145/1597036.1597038}
}

@inproceedings{AcharyaDJOPS12,
  author       = {Jayadev Acharya and
                  Hirakendu Das and
                  Ashkan Jafarpour and
                  Alon Orlitsky and
                  Shengjun Pan and
                  Ananda Theertha Suresh},
  editor       = {Shie Mannor and
                  Nathan Srebro and
                  Robert C. Williamson},
  title        = {Competitive Classification and Closeness Testing},
  booktitle    = {{COLT} 2012 - The 25th Annual Conference on Learning Theory, June
                  25-27, 2012, Edinburgh, Scotland},
  series       = {{JMLR} Proceedings},
  volume       = {23},
  pages        = {22.1--22.18},
  publisher    = {JMLR.org},
  year         = {2012},
  url          = {http://proceedings.mlr.press/v23/acharya12/acharya12.pdf}
}

@article{ValiantV17,
  author       = {Gregory Valiant and
                  Paul Valiant},
  title        = {Estimating the Unseen: Improved Estimators for Entropy and Other Properties},
  journal      = {J. {ACM}},
  volume       = {64},
  number       = {6},
  pages        = {37:1--37:41},
  year         = {2017},
  url          = {https://doi.org/10.1145/3125643},
  doi          = {10.1145/3125643},
  bibsource    = {dblp computer science bibliography, https://dblp.org}
}

@article{RaskhodnikovaRSS09,
  author       = {Sofya Raskhodnikova and
                  Dana Ron and
                  Amir Shpilka and
                  Adam D. Smith},
  title        = {Strong Lower Bounds for Approximating Distribution Support Size and
                  the Distinct Elements Problem},
  journal      = {{SIAM} J. Comput.},
  volume       = {39},
  number       = {3},
  pages        = {813--842},
  year         = {2009},
  url          = {https://doi.org/10.1137/070701649},
  doi          = {10.1137/070701649}
}

@article{BatuDKR05,
  author       = {Tugkan Batu and
                  Sanjoy Dasgupta and
                  Ravi Kumar and
                  Ronitt Rubinfeld},
  title        = {The Complexity of Approximating the Entropy},
  journal      = {{SIAM} J. Comput.},
  volume       = {35},
  number       = {1},
  pages        = {132--150},
  year         = {2005},
  url          = {https://doi.org/10.1137/S0097539702403645},
  doi          = {10.1137/S0097539702403645}
}

@article{FortnowRS94,
  author       = {Lance Fortnow and
                  John Rompel and
                  Michael Sipser},
  title        = {On the Power of Multi-Prover Interactive Protocols},
  journal      = {Theor. Comput. Sci.},
  volume       = {134},
  number       = {2},
  pages        = {545--557},
  year         = {1994}
}

@inproceedings{CanettiRR11,
  author       = {Ran Canetti and
                  Ben Riva and
                  Guy N. Rothblum},
  editor       = {Yan Chen and
                  George Danezis and
                  Vitaly Shmatikov},
  title        = {Practical delegation of computation using multiple servers},
  booktitle    = {Proceedings of the 18th {ACM} Conference on Computer and Communications
                  Security, {CCS} 2011, Chicago, Illinois, USA, October 17-21, 2011},
  pages        = {445--454},
  publisher    = {{ACM}},
  year         = {2011},
  doi          = {10.1145/2046707.2046759},
}

@article{CanettiRR13,
  author       = {Ran Canetti and
                  Ben Riva and
                  Guy N. Rothblum},
  title        = {Refereed delegation of computation},
  journal      = {Inf. Comput.},
  volume       = {226},
  pages        = {16--36},
  year         = {2013},
  doi          = {10.1016/J.IC.2013.03.003},
}

@article{GoldwasserKR15,
  author       = {Shafi Goldwasser and
                  Yael Tauman Kalai and
                  Guy N. Rothblum},
  title        = {Delegating Computation: Interactive Proofs for Muggles},
  journal      = {J. {ACM}},
  volume       = {62},
  number       = {4},
  pages        = {27:1--27:64},
  year         = {2015},
  doi          = {10.1145/2699436},
}

@article{CanettiLW25,
    title = {Refereed {Learning}},
    url = {http://arxiv.org/abs/2510.05440},
    doi = {10.48550/arXiv.2510.05440},
    language = {en},
    urldate = {2025-10-22},
    publisher = {arXiv},
    author = {Canetti, Ran and Linder, Ephraim and Wagaman, Connor},
    year = {2025}
}

@inproceedings{CanonneR14,
  author       = {Cl{\'{e}}ment L. Canonne and
                  Ronitt Rubinfeld},
  editor       = {Javier Esparza and
                  Pierre Fraigniaud and
                  Thore Husfeldt and
                  Elias Koutsoupias},
  title        = {Testing Probability Distributions Underlying Aggregated Data},
  booktitle    = {Automata, Languages, and Programming - 41st International Colloquium,
                  {ICALP} 2014, Copenhagen, Denmark, July 8-11, 2014, Proceedings, Part
                  {I}},
  series       = {Lecture Notes in Computer Science},
  volume       = {8572},
  pages        = {283--295},
  publisher    = {Springer},
  year         = {2014},
  url          = {https://doi.org/10.1007/978-3-662-43948-7\_24},
  doi          = {10.1007/978-3-662-43948-7\_24}
}

@article{GoldwasserMR89,
  author       = {Shafi Goldwasser and
                  Silvio Micali and
                  Charles Rackoff},
  title        = {The Knowledge Complexity of Interactive Proof Systems},
  journal      = {{SIAM} J. Comput.},
  volume       = {18},
  number       = {1},
  pages        = {186--208},
  year         = {1989},
  doi          = {10.1137/0218012}
}

@article{GoldreichMW91,
  author       = {Oded Goldreich and
                  Silvio Micali and
                  Avi Wigderson},
  title        = {Proofs that Yield Nothing But Their Validity for All Languages in
                  {NP} Have Zero-Knowledge Proof Systems},
  journal      = {J. {ACM}},
  volume       = {38},
  number       = {3},
  pages        = {691--729},
  year         = {1991},
  doi          = {10.1145/116825.116852}
}

@incollection{Goldreich20,
  author       = {Oded Goldreich},
  title        = {The Uniform Distribution Is Complete with Respect to Testing Identity
                  to a Fixed Distribution},
  booktitle    = {Computational Complexity and Property Testing},
  series       = {Lecture Notes in Computer Science},
  volume       = {12050},
  pages        = {152--172},
  publisher    = {Springer},
  year         = {2020}
}

@book{Vadhan99,
    AUTHOR = {Vadhan, Salil Pravin},
     TITLE = {A study of statistical zero-knowledge proofs},
      NOTE = {Thesis (Ph.D.)--Massachusetts Institute of Technology},
 PUBLISHER = {ProQuest LLC, Ann Arbor, MI},
      YEAR = {1999},
   MRCLASS = {Thesis},
  MRNUMBER = {2716914},
}

@incollection{Goldreich25,
  author       = {Oded Goldreich},
  title        = {On Parallel Repetition of Interactive Proof Systems},
  booktitle    = {Computational Complexity and Local Algorithms},
  series       = {Lecture Notes in Computer Science},
  volume       = {15700},
  pages        = {119--126},
  publisher    = {Springer},
  year         = {2025}
}

@inproceedings{Cleve86,
  author    = {Richard Cleve},
  title     = {Limits on the Security of Coin Flips when Half the Processors Are Faulty},
  booktitle = {Proceedings of the 18th Annual ACM Symposium on Theory of Computing (STOC '86)},
  pages     = {364--369},
  publisher = {ACM},
  year      = {1986},
  doi       = {10.1145/12130.12168},
  isbn      = {0-89791-193-8}
}

@inproceedings{AndrychowiczSP14,
  author    = {Marcin Andrychowicz and Stefan Dziembowski and Daniel Malinowski and {\L}ukasz Mazurek},
  title     = {Secure Multiparty Computations on Bitcoin},
  booktitle = {2014 IEEE Symposium on Security and Privacy},
  pages     = {443--458},
  publisher = {IEEE},
  year      = {2014},
  doi       = {10.1109/SP.2014.35}
}

@inproceedings{BentovKumaresan14,
  author    = {Iddo Bentov and Ranjit Kumaresan},
  title     = {How to Use Bitcoin to Design Fair Protocols},
  booktitle = {Advances in Cryptology -- CRYPTO 2014},
  series    = {Lecture Notes in Computer Science},
  volume    = {8617},
  pages     = {421--439},
  publisher = {Springer},
  year      = {2014},
  editor    = {Juan A. Garay and Rosario Gennaro},
  doi       = {10.1007/978-3-662-44381-1_24}
}

@article{Back14,
  author    = {Adam Back and Iddo Bentov},
  title     = {Note on fair coin toss via {Bitcoin}},
  journal   = {CoRR},
  volume    = {abs/1402.3698},
  year      = {2014},
  url       = {http://arxiv.org/abs/1402.3698},
  eprinttype= {arXiv},
  eprint    = {1402.3698}
}

@inproceedings{BaumEtAl20,
  author    = {Carsten Baum and Bernardo David and Rafael Dowsley},
  title     = {Insured MPC: Efficient Secure Computation with Financial Penalties},
  booktitle = {Financial Cryptography and Data Security (FC 2020)},
  series    = {Lecture Notes in Computer Science},
  volume    = {12059},
  pages     = {404--420},
  publisher = {Springer},
  year      = {2020},
  editor    = {Joseph Bonneau and Nadia Heninger},
  doi       = {10.1007/978-3-030-51280-4_22}
}

@inproceedings{chung2018game,
  title={Game theoretic notions of fairness in multi-party coin toss},
  author={Chung, Kai-Min and Guo, Yue and Lin, Wei-Kai and Pass, Rafael and Shi, Elaine},
  booktitle={Theory of Cryptography Conference},
  pages={563--596},
  year={2018},
  organization={Springer}
}

@inproceedings{thyagarajan2024game,
  title={Game-Theoretically Fair Distributed Sampling},
  author={Thyagarajan, Sri AravindaKrishnan and Soni, Pratik and Wu, Ke},
  booktitle={Annual International Cryptology Conference},
  pages={207--239},
  year={2024},
  organization={Springer}
}

@inproceedings{wu2022complete,
  title={A complete characterization of game-theoretically fair, multi-party coin toss},
  author={Wu, Ke and Asharov, Gilad and Shi, Elaine},
  booktitle={Annual International Conference on the Theory and Applications of Cryptographic Techniques},
  pages={120--149},
  year={2022},
  organization={Springer}
}

@inproceedings{AcharyaDK15,
  author       = {Jayadev Acharya and
                  Constantinos Daskalakis and
                  Gautam Kamath},
  title        = {Optimal Testing for Properties of Distributions},
  booktitle    = {{NIPS}},
  pages        = {3591--3599},
  year         = {2015}
}

@inproceedings{DiakonikolasK16,
  author       = {Ilias Diakonikolas and
                  Daniel M. Kane},
  title        = {A New Approach for Testing Properties of Discrete Distributions},
  booktitle    = {{FOCS}},
  pages        = {685--694},
  publisher    = {{IEEE} Computer Society},
  year         = {2016}
}

@article{LeviRR13,
  author       = {Reut Levi and
                  Dana Ron and
                  Ronitt Rubinfeld},
  title        = {Testing Properties of Collections of Distributions},
  journal      = {Theory Comput.},
  volume       = {9},
  pages        = {295--347},
  year         = {2013}
}

@inproceedings{LeviRR12,
  author       = {Reut Levi and
                  Dana Ron and
                  Ronitt Rubinfeld},
  title        = {Testing Similar Means},
  booktitle    = {{ICALP} {(1)}},
  series       = {Lecture Notes in Computer Science},
  volume       = {7391},
  pages        = {629--640},
  publisher    = {Springer},
  year         = {2012}
}

@inproceedings{BatuFFKRW01,
  author       = {Tugkan Batu and
                  Lance Fortnow and
                  Eldar Fischer and
                  Ravi Kumar and
                  Ronitt Rubinfeld and
                  Patrick White},
  title        = {Testing Random Variables for Independence and Identity},
  booktitle    = {{FOCS}},
  pages        = {442--451},
  publisher    = {{IEEE} Computer Society},
  year         = {2001}
}

@misc{Prashant21Lec03,
  author       = {Prashant Nalini Vasudevan},
  title        = {{L}ecture 03: Resources in Interactive Proofs},
  booktitle    = {CS6230: {T}opics in {I}nformation {S}ecurity},
  year         = {2021}
}

@article{ParnasRR06,
  author       = {Michal Parnas and
                  Dana Ron and
                  Ronitt Rubinfeld},
  title        = {Tolerant property testing and distance approximation},
  journal      = {J. Comput. Syst. Sci.},
  volume       = {72},
  number       = {6},
  pages        = {1012--1042},
  year         = {2006}
}

@inproceedings{CanonneJKL22,
  author       = {Cl{\'{e}}ment L. Canonne and
                  Ayush Jain and
                  Gautam Kamath and
                  Jerry Li},
  title        = {The Price of Tolerance in Distribution Testing},
  booktitle    = {{COLT}},
  series       = {Proceedings of Machine Learning Research},
  volume       = {178},
  pages        = {573--624},
  publisher    = {{PMLR}},
  year         = {2022}
}

@article {WuY16,
    AUTHOR = {Wu, Yihong and Yang, Pengkun},
     TITLE = {Minimax rates of entropy estimation on large alphabets via
              best polynomial approximation},
   JOURNAL = {IEEE Trans. Inform. Theory},
  FJOURNAL = {Institute of Electrical and Electronics Engineers.
              Transactions on Information Theory},
    VOLUME = {62},
      YEAR = {2016},
    NUMBER = {6},
     PAGES = {3702--3720},
      ISSN = {0018-9448},
   MRCLASS = {94A17},
  MRNUMBER = {3506758},
       DOI = {10.1109/TIT.2016.2548468},
       URL = {https://doi.org/10.1109/TIT.2016.2548468},
}

@inproceedings{ChanDSS14,
  author       = {Siu On Chan and
                  Ilias Diakonikolas and
                  Rocco A. Servedio and
                  Xiaorui Sun},
  title        = {Efficient density estimation via piecewise polynomial approximation},
  booktitle    = {{STOC}},
  pages        = {604--613},
  publisher    = {{ACM}},
  year         = {2014}
}

@inproceedings{ChanDSS13,
  author       = {Siu On Chan and
                  Ilias Diakonikolas and
                  Rocco A. Servedio and
                  Xiaorui Sun},
  title        = {Learning mixtures of structured distributions over discrete domains},
  booktitle    = {{SODA}},
  pages        = {1380--1394},
  publisher    = {{SIAM}},
  year         = {2013}
}

@article{DaskalakisDS15,
  author       = {Constantinos Daskalakis and
                  Ilias Diakonikolas and
                  Rocco A. Servedio},
  title        = {Learning Poisson Binomial Distributions},
  journal      = {Algorithmica},
  volume       = {72},
  number       = {1},
  pages        = {316--357},
  year         = {2015}
}

@inproceedings{DaskalakisDOST13,
  author       = {Constantinos Daskalakis and
                  Ilias Diakonikolas and
                  Ryan O'Donnell and
                  Rocco A. Servedio and
                  Li{-}Yang Tan},
  title        = {Learning Sums of Independent Integer Random Variables},
  booktitle    = {{FOCS}},
  pages        = {217--226},
  publisher    = {{IEEE} Computer Society},
  year         = {2013}
}

@article{Lindell20,
author = {Lindell, Yehuda},
title = {Secure multiparty computation},
year = {2020},
issue_date = {January 2021},
publisher = {Association for Computing Machinery},
address = {New York, NY, USA},
volume = {64},
number = {1},
issn = {0001-0782},
url = {https://doi.org/10.1145/3387108},
doi = {10.1145/3387108},
journal = {Commun. ACM},
month = dec,
pages = {86–96},
numpages = {11}
}

@inproceedings{OnakS18,
  author       = {Krzysztof Onak and
                  Xiaorui Sun},
  title        = {Probability-Revealing Samples},
  booktitle    = {{AISTATS}},
  series       = {Proceedings of Machine Learning Research},
  volume       = {84},
  pages        = {2018--2026},
  publisher    = {{PMLR}},
  year         = {2018}
}

\appendix
\section{Recap: Proof sketch of the Polarization Lemma (Lemma~\ref{lemma:polarization})}
    \label{sec:appendix}
We recapitulate the proof of polarization lemma in \cite{Vadhan99}.  In what follows, $\ell,m,r \geq 1$ are parameters to be optimized. Note that one can simulate a distribution resulting from the XOR Lemma (\cite[Lemma 3.1.16]{Vadhan99}, with parameter $\ell$) and the Direct Product Lemma (Lemma 3.1.15 in \cite{Vadhan99}, with parameter $m$) using $\ell$ and $m$ samples respectively.
    In what follows, we assume without loss of generality that $\errprob \in (0,c]$ for a sufficiently small absolute constant $c>0$ ($c=1/8$ suffices), as one can replace $\errprob$ by $\min(\errprob,c)$ at the cost of only constant factors in the claimed result. 
    \begin{itemize}
        \item Apply the XOR lemma on $\p,\q$ $\ell$ times to get $\p''',\q'''$. Then
        \begin{align*}
            \totalvardist{\p}{\q} \geq \alpha &\implies \totalvardist{\p'''}{\q'''} \geq \alpha^\ell \\
            \totalvardist{\p}{\q} \leq \beta &\implies \totalvardist{\p'''}{\q'''} \leq \beta^\ell
        \end{align*}
        \item Apply the Direct Product lemma on $\p''',\q'''$ $m$ times to get $\p'',\q''$. Then 
        \begin{align*}
            \totalvardist{\p'''}{\q'''} \geq \alpha^\ell &\implies \totalvardist{\p''}{\q''} \geq 1-2e^{-\frac{m\alpha^{2\ell}}{2}} \\
            \totalvardist{\p'''}{\q'''} \leq \beta^\ell &\implies \totalvardist{\p''}{\q''} \leq m \beta^\ell
        \end{align*}
        \item Finally, apply the XOR lemma $r$ times on $\p'',\q''$ to get $\p',\q'$. Then
        \begin{align*}
            \totalvardist{\p''}{\q''} \geq 1-2e^{-\frac{-m\alpha^{2\ell}}{2}} &\implies \totalvardist{\p'}{\q'} \geq \Paren{ 1-2e^{\frac{m\alpha^{2\ell}}{2}} }^r \geq 1-2r e^{-\frac{m\alpha^{2\ell}}{2}}\\
            \totalvardist{\p''}{\q''} \leq m \beta^\ell &\implies \totalvardist{\p'}{\q'} \leq \left( m \beta^\ell \right)^r
        \end{align*}
    \end{itemize}
    Overall, generating a sample from $\p'$ or $\q'$ takes $s\eqdef \ell m r$ samples from $\p,\q$. To get a concrete bound, set 
    \[
    t = 2 \ln \frac{1}{\errprob},\; \lambda \eqdef \min \left\{
    \frac{\alpha^2}{\beta}, 2 \right\} \in(1,2],\; 
    \ell \eqdef \clg{\log_{\lambda} 4 t},
    m \eqdef \frac{\lambda^\ell}{2 \alpha^{2 \ell}} \leq \frac{1}{2 \beta^\ell},\;
    r \eqdef    \clg{\log \frac{1}{\errprob}};
    \] 
    one can verify that
    $(m \beta^\ell)^r \leq \left( \frac{1}{2} \right)^r \leq \errprob$, and that 
    \[ 2 r e^{- \frac{m \alpha^{2 \ell}}{2}} = 2 r \exp \left( - \frac{\lambda^\ell}{4}
       \right) \leq 2 r \exp \left( - \frac{\lambda^{\log_{\lambda} 4 t}}{4}
       \right) = 2 r \exp (- t) = 2 \left\lceil \log \frac{1}{\errprob} \right\rceil
       \cdot \errprob^2 \leq \errprob ; 
    \]
    (the last inequality as $\errprob \leq c$); 
    and since
    \begin{eqnarray*}
      m = \frac{\lambda^\ell }{ 2 \alpha^{2 \ell}} & \leq & \frac{1}{2} 
      \frac{\lambda^{\log_{\lambda}(4 t) + 1}}{\alpha^{2 (\log_{\lambda}(4 t) +
      1)}}\\
      & = & \frac{1}{2}  \frac{4 t \lambda}{\alpha^2 \cdot \alpha^{\log_{\alpha}
      4 t \cdot 2 \frac{\log \alpha}{\log \lambda}}}\\
      & \leq & \frac{1}{2}  \frac{4 t \lambda}{\alpha^2 \cdot (4
      t)^{\frac{2\log \alpha}{\log \lambda}}}\\
      & = & \frac{\lambda}{2 \alpha^2}  (4 t)^{1 - \frac{2 \log \alpha}{\log
      \lambda}}\\
      & \leq & \frac{\lambda}{2 \alpha^2}  (4 t)^{1 + 2 \log (1 / \alpha)\cdot \max\Paren{\frac{1}{\log (\alpha^2 / \beta)},1}}\\
      & \leq & \min \left( \frac{1}{2 \beta}, \frac{1}{\alpha^2} \right)  (4
      \log (1 / \errprob))^{1 + \frac{2 \log (1 / \alpha)}{\log (\alpha^2 /
       \beta)}+ 2 \log (1 / \alpha)}
      .
    \end{eqnarray*}
    and so, in particular,
    \[ s = \ell m r \leq 
       (\log (1 / \errprob))^{\frac{2 \log (1 / \alpha)}{\log (\alpha^2 /
       \beta)}+ O(\log (1 / \alpha))} .
       \]
    This concludes the proof. \qed

\end{document}